\documentclass[12pt,preprint]{aastex}
\usepackage{amsmath,amssymb}

\newcommand{\project}[1]{\textsl{#1}}
\newcommand{\TheCannon}{\project{The~Cannon}}

\newcommand{\acronym}[1]{{\small{#1}}}

\newcommand{\sdssiii}{\project{\acronym{SDSS-III}}}
\newcommand{\apogee}{\project{\acronym{APOGEE}}}
\newcommand{\aspcap}{\project{\acronym{ASPCAP}}}
\newcommand{\lasso}{\project{\acronym{LASSO}}}
\newcommand{\dr}{\acronym{DR12}}
\newcommand{\logg}{\log g}
\newcommand{\mh}{\mathrm{[M/H]}}
\newcommand{\Teff}{T_{\mathrm{eff}}}
\newcommand{\Dvector}[1]{\boldsymbol{#1}}
\newcommand{\vectheta}{\Dvector{\theta}}
\newcommand{\vecv}{\Dvector{v}}
\newcommand{\argmin}[1]{\underset{#1}{\operatorname{argmin}}\,}
\newcommand{\pn}{\hphantom{-}}

\begin{document}

\title{\textsl{The Cannon 2:} A data-driven model of stellar spectra \\
       for detailed chemical abundance analyses}
\author{Andrew~R.~Casey\altaffilmark{1},
        David~W.~Hogg\altaffilmark{2,3,4,5},
        Melissa~Ness\altaffilmark{5},\\
        Hans-Walter~Rix\altaffilmark{5},
        Anna~Y.~Q.~Ho\altaffilmark{6},
    and~Gerry~Gilmore\altaffilmark{1}}
\altaffiltext{1}{Institute of Astronomy, University of Cambridge, 
                 Madingley Road, Cambridge CB3~0HA, UK}    
\altaffiltext{2}{Simons Center for Data Analysis, 160 Fifth Avenue, 7th Floor, 
                 New York, NY 10010, USA}
\altaffiltext{3}{Center for Cosmology and Particle Physics, 
                 Department of Physics, New York University, 4 Washington Pl., 
                 room 424, New York, NY, 10003, USA}
\altaffiltext{4}{Center for Data Science, New York University, 726 Broadway, 
                 7th Floor, New York, NY 10003, USA}
\altaffiltext{5}{Max-Planck-Institut f\"ur Astronomie, K\"onigstuhl 17, 
                 D-69117 Heidelberg, Germany}
\altaffiltext{6}{Astronomy Department, California Institute of Technology,
                 MC 249-17, 1200 East California Blvd, Pasadena, CA 91125, USA}
\email{arc@ast.cam.ac.uk}

\begin{abstract}
We have shown that data-driven models are effective for inferring physical 
attributes of stars (labels; $\Teff$, $\logg$, $\mh$) from spectra, even when
the signal-to-noise ratio is low.
Here we explore whether this is possible when the dimensionality of the label
space is large ($\Teff$, $\logg$, and 15 abundances: C, N, O, Na, Mg, Al, Si, S, 
K, Ca, Ti, V, Mn, Fe, Ni) and the model is non-linear in its response to 
abundance and parameter changes.
We adopt ideas from compressed sensing to limit overall model complexity
while retaining model freedom.  The model is trained
with a set of 12,681 red giant stars with high signal-to-noise spectroscopic 
observations and stellar parameters and abundances taken from the \apogee\ 
Survey.
We find that we can successfully train and use a model with 17 stellar labels.
Validation shows that the model does a good job of inferring all 17 labels 
(typical abundance precision is 0.04~dex), even when we degrade the signal-to-noise by 
discarding $\gtrsim$50\% of the observing time. The model dependencies 
make sense: the spectral derivatives with respect to abundances correlate
with known atomic lines, and we identify elements belonging
to atomic lines that were previously unknown.  We recover (anti-)correlations
in abundance labels for globular cluster stars, consistent with the literature.
However we find that the intrinsic spread in globular cluster abundances is 3--4 times smaller than 
previously reported.  We deliver 17 labels with associated errors for 
87,563 red giant stars, as well as open-source code to extend this work to other spectroscopic surveys.
\end{abstract}

\section{Introduction}
The detailed chemical composition of a star's photosphere reflects its
 formation environment, and the mix of supernovae that 
preceded it.  This photospheric \emph{chemical fingerprint} remains largely
unchanged throughout a star's lifetime, providing a fossil record of 
local star-formation history.  Although the differences in detailed chemical
abundances may be subtle between two formation sites, if stars could be linked
to their natal gas cloud by their chemical fingerprint, a sufficiently large
collection of precise stellar abundances would unravel the
complete chemical evolution of the Milky Way.

This goal has only recently become feasible with the increasing volume and 
quality of stellar spectra obtained in the last decade.  Specifically, large 
surveys are obtaining high-resolution ($\mathcal{R} \gtrsim 20,000$), high 
signal-to-noise (S/N) ratio spectra for $\sim10^5$---$10^6$ stars across all 
components of the Galaxy \citep{Gilmore_2012,Zasowski_2013,De_Silva_2015}.  This is a sharp relative increase in data volume; it has mandated the automation of spectral analysis, and encouraged dozens of
groups to produce bespoke pipelines.

Most automated pipelines have grown from classical, manual methods.  There has 
been relatively little work on unconventional methods to analyse spectra.  
Spectroscopists have instead sought to code their experience (or subjectivity), 
with heuristics enforced for wavelength masks, convergence criteria, 
and similar analysis issues.  Most of these decisions are based on the optimization of accuracy for particular stars with good quality data (e.g., Solar-like
stars with high S/N ratios).  As a consequence, these heuristics are frequently 
incompatible for data with more modest (and representative) S/N ratios.  Indeed, 
it can be shown from repeat or blind experiments that traditional pipelines 
routinely yield imprecise abundances for noisy data.  Moreover, the results from individual 
pipelines are inconsistent, with differences an order of 
magnitude larger than what state-of-the-art spectroscopic studies seek to 
measure (e.g., the effects of atomic diffusion, evolutionary mixing, planetary
accretion).  Thus, while there has been substantial effort to automate
classical analysis techniques, they are generally imprecise at modest S/N 
ratios, and frequently yield inaccurate (biased) results.

Considerable effort has  been spent on improving the accuracy of physical
models of stars.  Indeed it is impossible to measure physical properties of stars (or 
their chemical abundances) accurately without accurate physical models of 
stellar spectra.  However physics-based models of stars have a number of known
problems: they rely on incorrect approximations (e.g. one-dimensional atmospheres) and incomplete laboratory data.
Three-dimensional models remain computationally impractical for more than just a
few stars.  As a consequence of the limited atmosphere dimensionality, crude 
(knowingly incorrect) approximations for the convection are 
necessary.  Although some grids of three-dimensional hydrodynamic models have been 
produced and averaged to one-dimensional approximations, these models assume 
local thermodynamic equilibrium (LTE).  Properly accounting for departures from 
LTE is a formidable analytic and computational challenge.  Additionally, while 
laboratory efforts have thoroughly improved much of the faulty atomic and 
molecular data, this process is unquestionably incomplete.  For these reasons 
there are some spectral features that are much better understood than others.  
As a consequence, physics-based methods are restricted to spectral regions and
parameter spaces that are understood marginally better, which limits 
their applicability and interpretability \emph{by construction}.

In detail, physics-based models do not explain all pixels of a stellar 
spectrum at the precision with which we are currently observing.  The data
quality have outgrown the classical methods used to analyse them.  This led to the creation of
\TheCannon\footnote{It is important (to us) to note that \TheCannon\ is named 
not after a weapon but instead after Annie Jump Cannon, who was the first to 
correctly order stellar spectra in temperature order \citep[and who did
so by looking at the data, and without any use of physics-based models, see, e.g.,][]
{Cannon_1912}.} (\citealt{tc}), a data-driven---as opposed to 
physics-based---model for stellar spectra.

Before we continue to introduce background on this work, we first need to
introduce some relevant terminology.  Throughout this work we will call stellar 
parameters ($\Teff$ and $\logg$) and the full set of 15 chemical abundances 
collectively ``labels''.  This unifies and collapses the phrase ``stellar
parameters and chemical abundances'' to a word, and connects to relevant 
terminology for supervised methods in the machine learning and statistics 
literatures.  \TheCannon\ uses spectra from stars that have labels known
with high fidelity to train a model for stellar spectra.  The trained model can
then be used on new data, and precisely estimate labels.

  \TheCannon\ is a data-driven model, 
but it differs from standard machine learning approaches because it contains an 
explicit noise model.  This means that \TheCannon\ can transfer labels from high
S/N training set stars to low S/N test set stars; that is, the training set and
the test set do not need to be statistically identical.  This is related to the
fact that \TheCannon\ is an interpretable causal model; the internals of the model are 
the dependencies of the spectral expectation and variance on wavelength and 
physical parameters of the star, plus an explicit noise model.  Given a representative set of stars with known
labels of high-fidelity, \TheCannon\ provides a generative model for stellar 
spectra based on a non-linear combination of the labels.

While a physical model is required to construct an acceptable training set for
\TheCannon\ to use, \TheCannon\ also offers opportunities to reconcile known problems with physics-based 
models.  It provides labels for training set data at wavelengths where there is
limited or missing atomic information, allowing for every pixel to contribute in 
measuring labels from noisy spectra.  If there are a sufficient
number of stars observed by two surveys, \TheCannon\ can also be used to determine labels that
are consistent across surveys and wavelengths \citep{Ho_2016}.  Moreover, as we
will show in this work, the model internals have physical interpretations, 
allowing for the identification of previously unknown spectral lines.  This is
a first step towards using \TheCannon\ to improve physics-based models.

Only a small number of labels were used in the first work with \TheCannon: 
three in the original work, and four or five in later work \citep{tc, age, Ho_2016}. 
Here we were guided by thoughts related to density estimation. To sample a 
$K$-dimensional label space well, the size of the training set should scale
exponentially (or worse) with $K$.  Subsequent experiments, however, did not 
bear this out.  We found that we can transfer many labels from the training set 
to the test set, with training sets of (just) thousands of stars.  The fundamental 
reason is that \TheCannon\ is \emph{not} a density estimator.  It is more like 
an \emph{interpolator}, which effectively finds stars in the training set that
are close to the test star, and transfers labels, using a polynomial 
model as a smooth interpolation function.

The capacity to extend to a larger set of labels without significant
computational detriment offers tantalizing opportunities.  The most 
straightforward is that it provides a measurement of many elemental abundances.
In doing so, however, it can be shown that a standard \emph{Cannon} model yields
noisy coefficients (spectral derivatives; see next Section) that are incompatible with 
expectations from physics: the training coefficients may optimize to produce 
coefficients of abundance labels with non-zero 
contributions at \emph{all pixels}, however physically we know that spectral lines of most
elements do not contribute at all wavelengths.

For this reason alone we know that the problem ought to be sparse.  Here we exploit this 
knowledge to the fullest, using standard regularization methods to discover
and enforce sparsity.  We consider the \emph{entire} 17-dimensional label space 
produced by the \apogee\ \aspcap\ pipeline.  For our training set, we adopt the \aspcap\ labels of the stars with 
the highest S/N spectra.  We show by validation that we can transfer these labels to much 
lower S/N stars, with reduced precision but no strong biases.  We then use the
system to label all of the stars in the \apogee\ \dr\ data set.  After
validating our model, we confirm our abundance precision using tests of globular
clusters.

\section{Method}

Before we outline our assumptions, we need to define the different data sets we
will use, as well as terminology related to our method. Consider a stellar spectroscopic
survey.  Within the survey data is a set of stars observed at high S/N, where the labels for those stars are known with high
fidelity.  We call this sample the \emph{reference set}, and all other stars
are set aside into the \emph{test set}.  Note that there may be labels
reported for stars in the \emph{test set}, but the implicit assumption is
that those labels are not known with high fidelity.  We are going to construct
(train) \TheCannon\ using a subset of the reference set, allowing us to 
\emph{predict} stellar fluxes and \emph{measure} (test) labels for stars in the 
test set.  It is important to note that we do not use every star in the
reference set for training: here we will define the \emph{training set} as a random
subset of the reference set.  All stars in the reference set that are not part of 
the training set will form the \emph{validation set}, which we will use to 
\emph{validate} the predictive power of our model.

\noindent{}We assume the following about \TheCannon\ and the \apogee\ \dr:
\begin{itemize}
\item
Stars with similar labels ($\Teff$, $\logg$, and abundances) have similar spectra.
\item
The expectation that the spectrum of a star is a smooth function of the values of 
the labels for that star.  Further than this, we assume that the function is so 
smooth it can be reasonably approximated with a quadratic form in label space.
\item
The resolution of all \apogee\ spectra are identical and all spectra are calibrated
to the same rest wavelength grid.
\item
The \apogee\ noise variances reported are (nearly) correct, normally distributed,
and independent from pixel to pixel.  
Importantly, we are \emph{not} assuming that different stars have similar noise
variances, nor that the reference and test sets have the same noise model.
\item
We have a training set of stars with \emph{mean} accurate labels, where ``accurate'' is 
defined by the accuracy requirements of the output labels.  It might be more 
appropriate to say that we are assuming that the training set stars have 
\emph{consistent} labels (consistent with the assumptions of smoothness, above).
Indeed, because \TheCannon\ is a data-driven model we stress that it cannot
create ``ground truth'': we are propagating labels from the training set.
\item
The training set is representative, in the sense that the training set stars 
span the label space similarly to how the test set spans the label
space.
\item
We assume that our continuum normalization procedure (described below) is consistent.
We do not require ``true'' continuum-normalization in the classical sense 
because any offset (even a label-dependent residual due to a strong absorption
line) can be captured by the model.  Instead we require that our normalization
procedure is invariant with respect to S/N.
\end{itemize}

\noindent{}Given these assumptions, the model we adopt is
\begin{eqnarray}
  y_{jn} &=& \vecv(\ell_n)\cdot\vectheta_j + e_{jn}
  \label{eq:model}\quad ,
\end{eqnarray}
where $y_{jn}$ is the data for star $n$ at wavelength pixel $j$, $\vecv(\ell_n)$
is a function that takes as input the label list $\ell_n$ of length $K$ for star
$n$ and outputs a vector of length $D>K$ of functions of those labels,
$\vectheta_j$ is a vector of length $D$ of parameters controlling the model at 
wavelength pixel $j$, and $e_{jn}$ is a noise draw or residual.  We refer to 
$\vecv(\ell_n)$ as ``the vectorizing function'', which allows for arbitrarily 
complex functions that might not be simple polynomial expansions of the label 
list $\ell_n$ (e.g., sums of sines and cosines).  Inasmuch as the model is good,
the noise values $e_{jn}$ can be taken to be drawn from a Gaussian with zero 
mean and variance $\sigma^2_{jn}+s^2_j$, where $\sigma^2_{jn}$ is the 
pipeline-reported uncertainty variance on datum $y_{jn}$ and $s^2_j$ is a 
parameter describing excess variance at wavelength pixel $j$.

Two comments about the model (\ref{eq:model}).  The first is that, because the 
$e_{jn}$ are thought of as being drawn from a probability density function (pdf),
it is a probabilistic model for the spectral data $y_{jn}$.  The second is that
the output of the function $\vecv(\ell)$ can be thought of as a row of the 
``design matrix'' that defines the possible freedom given to the spectrum 
expectation model.

In the \emph{training step}, we fix the $K$-lists of labels $\ell_n$ for all 
training set stars $n$.  We seek, at each wavelength pixel $j$, the $[D+1]$ 
parameters $\vectheta_j,s^2_j$ that optimize a penalized likelihood:
\begin{eqnarray}\label{eq:train}
  \vectheta_j,s^2_j &\leftarrow& \argmin{\vectheta,s}\left[
    \sum_{n=0}^{N-1} \frac{[y_{jn}-\vecv(\ell_n)\cdot\vectheta]^2}{\sigma^2_{jn}+s^2}
    + \sum_{n=0}^{N-1} \ln(\sigma^2_{jn}+s^2)
    + \Lambda_j\,Q(\vectheta)
    \right]
  \quad ,
\end{eqnarray}
where $\Lambda_j$ is a regularization parameter, and $Q(\vectheta)$ is a 
regularizing function that encourages parameters to take on zero values.  The 
regularizing function takes a $D$-vector as input and returns a scalar value.
We call this penalized likelihood---the argument of the argmin in 
equation~(\ref{eq:train})---the \emph{training scalar}.  We will adopt for the 
regularizing function $Q(\vectheta)$ in the training scalar a modification of L1
regularization, discussed below.  Although the training-step optimization 
problem will not in general be convex, we can make choices for $Q(\vectheta)$ 
(and we will) to make the problem such that it would be convex at any fixed 
value of $s^2$; for this reason it will tend to optimize well in most cases of 
interest. Convexity is extremely advantageous: it ensures that only a single
global minima exists, implying that any reasonable optimization routine is
guaranteed to optimize to the correct global minima (i.e., no local minima).

The regularization parameter $\Lambda_j$ sets the strength of the 
regularization; as $\Lambda_j$ increases, the number of non-zero components of 
the parameter vector $\vectheta_j$ will decrease.  We give the regularization 
parameter a subscript $j$ because in general we can set it differently at every
wavelength.  This makes sense, because different wavelengths have very different
dependences on components of the label list $\ell$.  In practice the
value of $\Lambda_j$ should be set by full cross-validation, and possibly 
include restrictions based on physical arguments (e.g., a particular element 
does not have any spectral lines near this pixel, therefore the contributions
from this label must be zero).  This vastly expands the number of potential 
hyper-parameters, and thus the computing expense required to determine them.  
Thus for the purpose of this work we will set a single value of $\Lambda$ (for 
all $j$ pixels) by validation.

In the \emph{test step}, we fix the parameters $\vectheta_j,s^2_j$ at all
wavelength pixels $j$.  We seek, for each test set star $m$, the $K$-list of 
labels $\ell_m$ that optimizes the likelihood:
\begin{eqnarray}\label{eq:test}
  \ell_m &\leftarrow& \argmin{\ell}\left[
    \sum_{j=0}^{J-1} \frac{[y_{jm}-\vecv(\ell)\cdot\vectheta_j]^2}{\sigma^2_{jm}+s^2_j}
    \right]
  \quad .
\end{eqnarray}
If the vectorizing function $\vecv(\ell)$ is non-linear (as it is in our 
quadratic model), the test-step optimization is not convex.  However, because
there are many pixels $j$ acting, each of which has a different functional
dependence on the labels in the label list $\ell$, in practice the optimization 
finds a good value for the label list $\ell$.  We call
this partial log likelihood---the argument of the argmin in 
equation~(\ref{eq:test})---the \emph{test scalar}.

The model freedom of \TheCannon\ is set by the vectorizing function 
$\vecv(\ell)$---which takes the $K$-element label list $\ell$ and expands it 
into a $D$-dimensional vector of components for the linear model---and the 
regularization $\Lambda_j\,Q(\vectheta)$.  Because we want the (simple, see 
below) regularization to treat the different parameters (the different 
components of $\vectheta$) in some sense ``equally'', we have to make sensible 
choices in the vectorizing function $\vecv(\ell)$.  One thing that the 
vectorizing function $\vecv(\ell)$ can do is offset the labels by some kind of
fiducial (mean, median, or other central) value, such that $\vecv(\ell)=0$ is at
a central location in the label space.  Another is to divide out a scale, 
because, for example, $\Teff$ values are in the thousands, but $\logg$ values
are of order unity.  If scale is not divided out, the (isotropic in $\vectheta$)
regularization will be much more harsh, effectively, on some parameters than 
others.  An extension of this work might be to consider different regularization
terms for each $j$ pixel \emph{and} every label in $\ell$.

In what follows, we adopt the median value in the training set for each label 
value as the fiducial offset.  We choose a dimensionless scale factor $f$ 
times the label range (defined as the difference between the 97.5th percentile
and the 2.5th percentile of the training set along each label direction) as 
the scale such that, for example, the $\Teff$ value for star $n$ is rescaled as:

\begin{eqnarray}\label{eq:label-norm}
  \hat{T}_{\mathrm{eff},n} = \frac{T_{{\rm eff},n} - p_{\Teff,50}}{f\cdot|p_{\Teff,97.5} - p_{\Teff,2.5}|}
\end{eqnarray}

\noindent{}where $p_{\theta,k}$ is the $k$-th percentile value of $\theta$ in
the training set labels. For the regularizer $Q(\vectheta)$ we adopt a variant
of L1 regularization; we set

\begin{eqnarray}\label{eq:l1-variant}
  Q(\vectheta) &=& \sum_{d=1}^{D-1} |\theta_d|
  \quad,
\end{eqnarray}
where the sum is over the $[D-1]$ components of $\vectheta$, excluding the 
zeroth component because we don't ever expect that component to vanish.\footnote{Forgive a notational similarity here: When we subscript (bold) $\vectheta$ 
with $j$ we mean ``the $D$-vector of parameters associated with wavelength pixel
$j$''; when we subscript (non-bold) $\theta$ with $d$ we mean ``the single parameter along 
coordinate axis $d$ in the $D$-dimensional $\vectheta$ vector space.''} There is
a great deal of theory about this kind of regularization; it is called L1 or the
\lasso\ \citep{Tibshirani_1996}.  L1 regularization encourages parameters to vanish 
precisely but doesn't break convexity for convex problems.  It is important to
note that the scale factor $f$ plays into the regularization, because as the 
scale factor grows, the more penalized the cross-terms (e.g., $\Teff\cdot$[Al/H], [Mg/H]$\cdot$[Si/H]) become 
relative to the linear terms.  We will heuristically set the values of the hyper-parameters
$\Lambda$ and $f$ by validation in Section \ref{sec:hyper-parameter-validation}.

\section{Training, validation, and test data}
\label{sec:training set}

We employ the \apogee\ Data Release 12 data \citep{Alam_2015} to demonstrate the effectiveness of
a regularized \emph{Cannon} model in high-dimensional label space.  We 
constructed the reference set using sensible criteria for stars analysed with 
version \texttt{v603} of the \aspcap\ pipeline \citep{Holtzman_2015,Garcia_Perez_2015}.  We first removed 
any stars with problematic flags from the \aspcap\ pipeline, i.e., we required 
\texttt{ASPCAPFLAG~=~0}.  We excluded stars with S/N ratios outside the range of
200-300, and stars with a radial velocity scatter larger than 1~km~s$^{-1}$.  We
further demanded that our training set include reported abundances in all 15 
elements (C, N, O, Na, Mg, Al, Si, S, K, Ca, Ti, V, Mn, Fe, and Ni), and 
restricted the abundance range such that $2 > \mathrm{[X/Fe]} > -2$, 
$\mathrm{[Fe/H]} > -3$, and $[\alpha/\mathrm{Fe}] > -0.1$.  A visual comparison 
of the [V/H] labels with other (Fe-peak) abundance labels showed that many stars
in the reference set had spurious measurements of [V/H].  For this reason we 
required that stars in the reference set have $\mathrm{[V/Fe]} > -0.6$.  The 
distilled sample includes 14,141 red giant branch stars that will form our 
reference set, with [Fe/H] ranging from $\mathrm{[Fe/H]} = -2.10$ to 
+0.30~dex.

\begin{figure}[p]
\centering
\includegraphics[width=\textwidth]{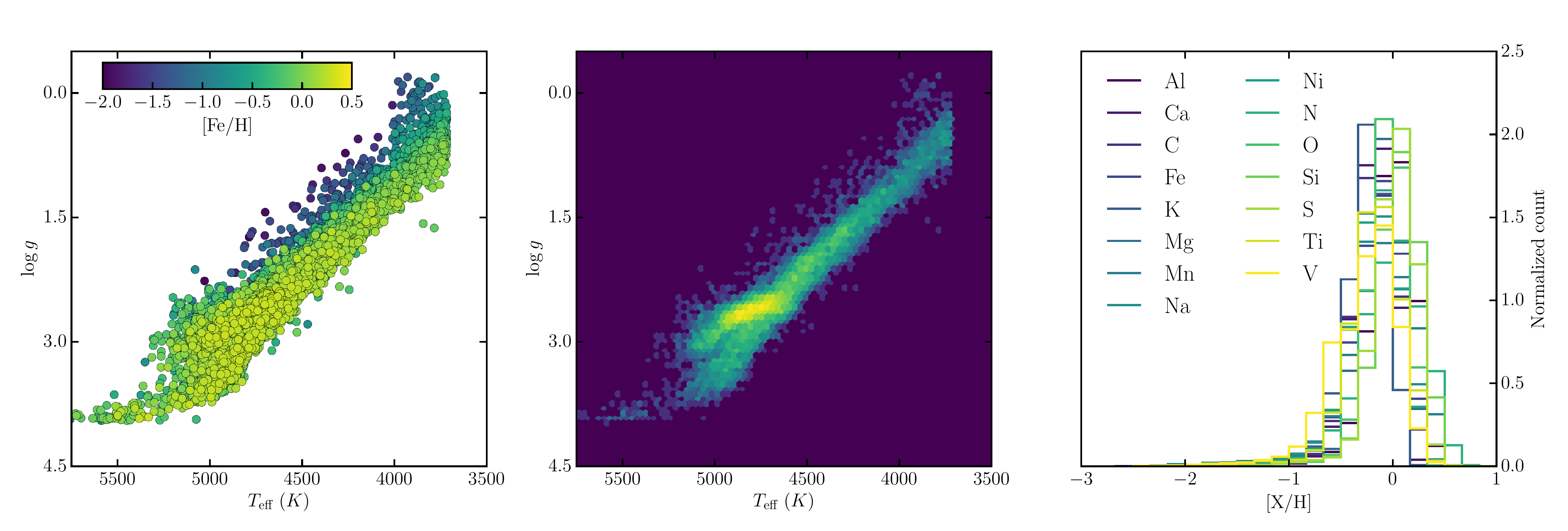}
\caption{A temperature-surface gravity diagram for all 12,681 stars in
the training set (left). A logarithmic density plot for the training set
is shown in the centre panel. Distributions of abundance labels
are shown in the right-hand panel.\label{fig:training_set_hrd}}
\end{figure}

We randomly assigned each star in the reference set a uniformly-distributed 
integer $q$ between 0 and 9, inclusive.  We assign stars with $q > 0$
 to the \emph{training set}, and those with $q = 0$ to the \emph{validation set}. We have ensured that the same $q$ integer was assigned to each star 
between the different experiments described below.  The training set (see Figure
\ref{fig:training_set_hrd}) comprises 12,681 red giant branch stars, and the 
validation set includes 1,460 stars.

Here we describe the requisite data processing steps before we can train the 
spectral model.  The \apogee\ \texttt{apStar} files contain rest-wavelength, 
resampled fluxes from individual visits for a given star, and an uncertainty array for
those fluxes.  Although the \texttt{aspcapStar} files contain stacked, 
pseudo-continuum normalized spectra for a given star, we chose to only use the 
fluxes from the \texttt{apStar} data files throughout this work.  Our reasoning 
is as follows: The \aspcap\ pipeline uses a running quartile filter window to determine 
the continuum.  For this reason it is provably variant with the S/N ratio (biased at low S/N): at 
low S/N ratios the inferred continuum will be systematically shifted with 
respect to the same star observed in high S/N.  Although our reference set only 
includes high S/N spectra, our results would suffer if we employed the \aspcap\ 
normalization procedure for low S/N spectra in the \emph{test set}.  Thus, we 
opted to normalize and stack the fluxes from individual visits provided in the 
\texttt{apStar} files. Our normalization procedure is described below.

Individual spectra in the \texttt{apStar} files contain associated uncertainty arrays
for each observation.  However the uncertainty arrays do not encapsulate all knowledge
about any technical, observational, or reduction issues.  Instead, every pixel
in a single observation also contains a bitmask flag that documents potential
issues.  Although pixels are flagged, these issues are not reflected in the
uncertainty arrays of individual observations.  For this reason we chose to construct
an (adjusted) inverse variance array that encapsulates the implied additional 
uncertainty in flagged pixels.  Specifically for pixel $j$, if it is flagged, we set the error
array such that:

\begin{equation}
\sigma_{j,adjusted}^2 = \sigma_{j}^2 + \Delta_{j}^2
\end{equation}

\noindent{}where

\begin{equation}
\Delta_{j} = \max{\left(C_{0}|y_{j} - \widetilde{y}|,C_{1}\widetilde{y}N_\mathrm{flagged}\right)} .
\label{eq:pixel_flag}
\end{equation}

Here $N_\mathrm{flagged}$ is the number of flagged pixels in the spectrum, and (what
we call the conservatism) constants $C_0$ and $C_1$ have been chosen as 
$\{2.0,0.1\}$, which produces sensible inverse variance values for flagged
pixels.  We chose to ignore bitmask values associated with persistence 
(specifically values 9, 10, and 11) as frequently every pixel in the blue CCD
would be flagged with these values, thereby producing unrealistically large 
uncertainties at every pixel in the blue CCD.  Consider the two limits of (\ref{eq:pixel_flag}):
pixel $A$ is flagged and has a flux value that is very far (say 5$\sigma$) 
away from the mean flux. Pixel $B$ is also flagged but has flux values very
similar to the mean spectrum -- a visual inspection suggests it is not anomalous.
For flagged pixels of the $A$ kind, the added variance is proportional to how
discrepant the flux is to the mean spectrum flux. On the other hand, there may be many 
$B$-like pixels that represent a subtle systematic problem. These pixels are
accounted for by considering the number of flagged pixels $N_\mathrm{flagged}$. In both
cases, our conservative approach accounts for underestimated variance in flagged pixels 
due to local ($A$-type) and global ($B$-type) problems.

After updating the inverse variance arrays to account for flagged pixels, we
pseduo-continuum-normalized the individual observations. The spectra were 
normalized in three different regions, corresponding to each CCD, between
wavelength regions 15090--15822~\AA, 15823--16451~\AA, and 16452--16971~\AA.
We distilled a list of continuum pixels \citep[following the initial 
identification in][]{tc}, and for each region in all observations we fit
the continuum-pixel fluxes as a sum of sine and cosines with $L = 1400$ and 
$W = 3$, solving for amplitudes $\Dvector{A}$ in:

\begin{equation}
y_j = \sum_{w=0}^{W} A_{2w}\sin{\left(\frac{2w\pi\lambda_{j}}{L}\right)} + A_{(2w+1)}\cos{\left(\frac{2w\pi\lambda_{j}}{L}\right)}.
\end{equation}

Using a sum of sines and cosines for pseudo-continuum normalization has a number
of advantages over alternative approaches.  Firstly, a design matrix can be
constructed \emph{a priori}, at which point the normalization procedure becomes a
linear matrix operation on fluxes and is therefore cheap.  More practically, 
the use of sine and cosine functions implies that edge behaviour of the 
continuum function will be bounded (more bounded than a polynomial expression),
and are therefore less susceptible to issues with overfitting where the edge of
the continuum function demonstrates large wiggles.

The approach we have described is \emph{pseudo}-continuum normalization.  While 
our choice of continuum pixels is well-informed \citep{tc}, our 
procedure does not require that the chosen pixels to are indeed ``true'' continuum
pixels.  This is advantageous.  The continuum pixels could be randomly selected; 
our approach would be equally effective as long as the same pixels were always
used for normalization.  All spectra would be ``normalized'' in the same way, and
the residual flux behaviour from a ``true''-normalized spectrum would be captured 
by the spectral model.  We have attempted to select well-informed continuum
pixels only in order to maximize model interpretability: by using reasonable 
continuum pixels we can be sure that the spectral model derivatives can be 
interpreted as being astrophysically motivated, rather than capturing residual 
continuum effects.

For all of the aforementioned reasons (invariance with respect to S/N, bounded
functions, linear operations, and repeatability between spectra), we emphasize
that if traditional continuum methods are employed, \TheCannon\ will (likely) 
give \emph{very bad results}.  Adopting a linear continuum normalization 
procedure is paramount.  After normalizing all individual spectra in the
\texttt{apStar} files, we re-stacked the spectra using the (adjusted) inverse
variance of each pixel as weights.  Therefore for all stars observed by \apogee,
we have normalized individual and combined spectra, allowing for a 
self-consistent examination of label determination at low S/N (see Section
\ref{sec:label-recovery-snr}).

\section{Experiments}
\label{sec:experiments}

\subsection{Hyper-parameter selection and validation}
\label{sec:hyper-parameter-validation}

In the simplest case we have two hyper-parameters that need to be determined: 
$\Lambda$ and $f$.  The strength of the regularization (at all pixels) is set 
by $\Lambda$, and the scale factor $f$ controls the scaling on individual 
labels.  In practice large $\Lambda$ values encourage zero coefficients and thus enforce sparsity.  High scale factor $f$
values act to penalize cross-terms more than linear labels, and produce sparse
models as a consequence.  Because these hyper-parameters act in conjunction 
with each other, we need to explore many combinations of $\Lambda$ and $f$.

Here we use the reference set described in the previous Section to perform a
grid search with different values of $\Lambda$ and $f$.  Specifically we vary 
$\Lambda$ in logarithmic steps from $10^0$ to $10^5$, and increase the 
scale factor $f$ from 0.5 to 50.  For each combination of $\Lambda$ and $f$ we 
trained the full 17-label model using the training set (the $\sim$90\% random
subset of the reference set) and measured labels for all spectra in the
validation set: both the combined and individual observations.  Given the model
complexity, for this experiment we fixed $s^2 = 0$ to leverage convexity
and to ensure convergence at the global minimum.  This also ensures a fair
comparison between different hyper-parameters, because if $s^2$ is free it
will inevitably rise at very high $\Lambda$ values.

We must chose heuristic(s) to select an appropriate regularization and scale
factor.  Many metrics are available to us: the predictive power in spectral
fluxes (e.g., the $\chi^2$ or test scalar for the validation set), the 
recovered precision in labels at low S/N, or something that balances goodness of fit and sparsity.

\begin{figure}[p]
\centering
\includegraphics[width=\textwidth]{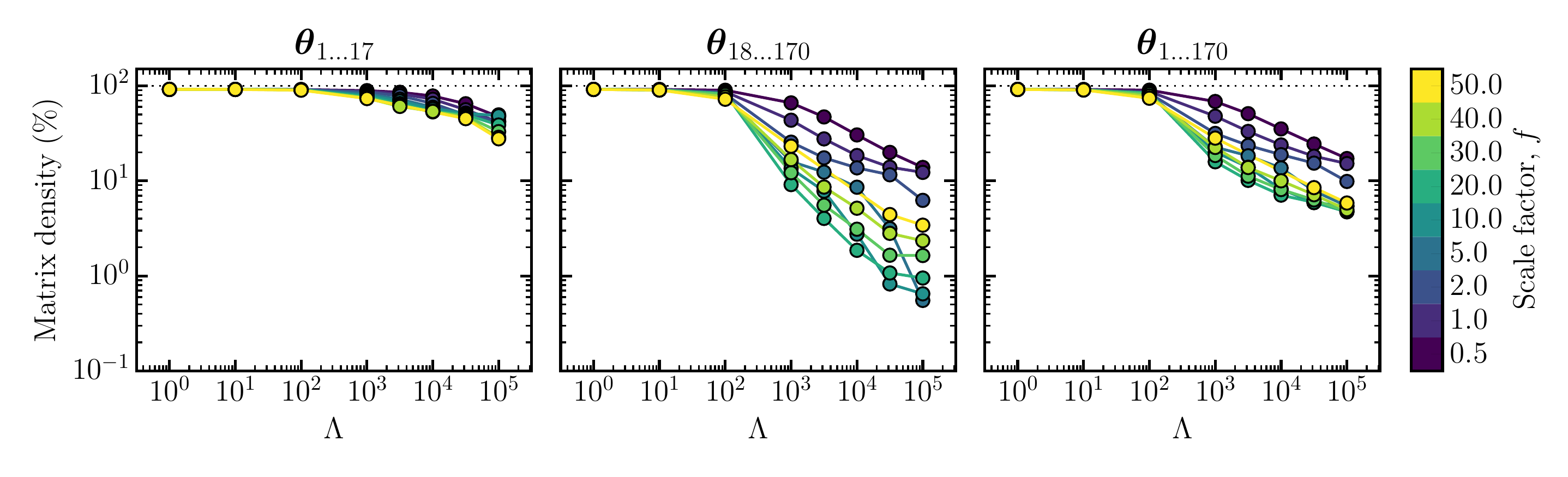}
\caption{The fraction of non-zero coefficients (i.e., a measure of matrix density) for 17-label models with different hyper-parameters $\Lambda$ and $f$.  The density of the first-order coefficients are shown in the left panel, the second-order coefficients (e.g., $\Teff^2$, $\Teff\cdot\logg$, or $[\rm{Al}/\rm{H}]\cdot[\rm{Mn}/\rm{H}]$) in the middle panel, and total density (excluding the baseline spectrum coefficient $\vectheta_0$) in the right-hand panel.\label{fig:sparsity}}
\end{figure}

We define the model sparsity as the percent of zero-value spectral derivatives 
$\vectheta$.  The sparsity could be calculated across the linear 
coefficients, only the cross-term coefficients, or some combination thereof.  
Note that we never include the baseline spectrum coefficients $\vectheta_0$ when 
calculating sparsity metrics because this parameter is not regularized (see 
Equation \ref{eq:l1-variant}).  In Figure \ref{fig:sparsity} we show three 
different sparsity metrics for many permutations of hyper-parameters $\Lambda$ 
and $f$.  Specifically we show the sparsity of the linear model coefficients 
$\vectheta_{1...17}$, the cross-term coefficients $\vectheta_{18...170}$, and 
the combination of linear and cross-term coefficients.  It is clear that the 
total model sparsity does not change significantly (regardless of $f$) until 
$\Lambda \gtrsim 10^3$.  In this regime the linear and second-order coefficient 
sparsity metrics exhibit very different responses.  The cross-term sparsity 
increases faster than the linear terms, with a clear (and expected) dependence 
on the scale factor $f$.  For any $\Lambda \gtrsim 10^3$, a scale factor 
$f \approx 20$ produces the sparsest model.

\begin{figure}[p]
\centering
\includegraphics[width=\textwidth]{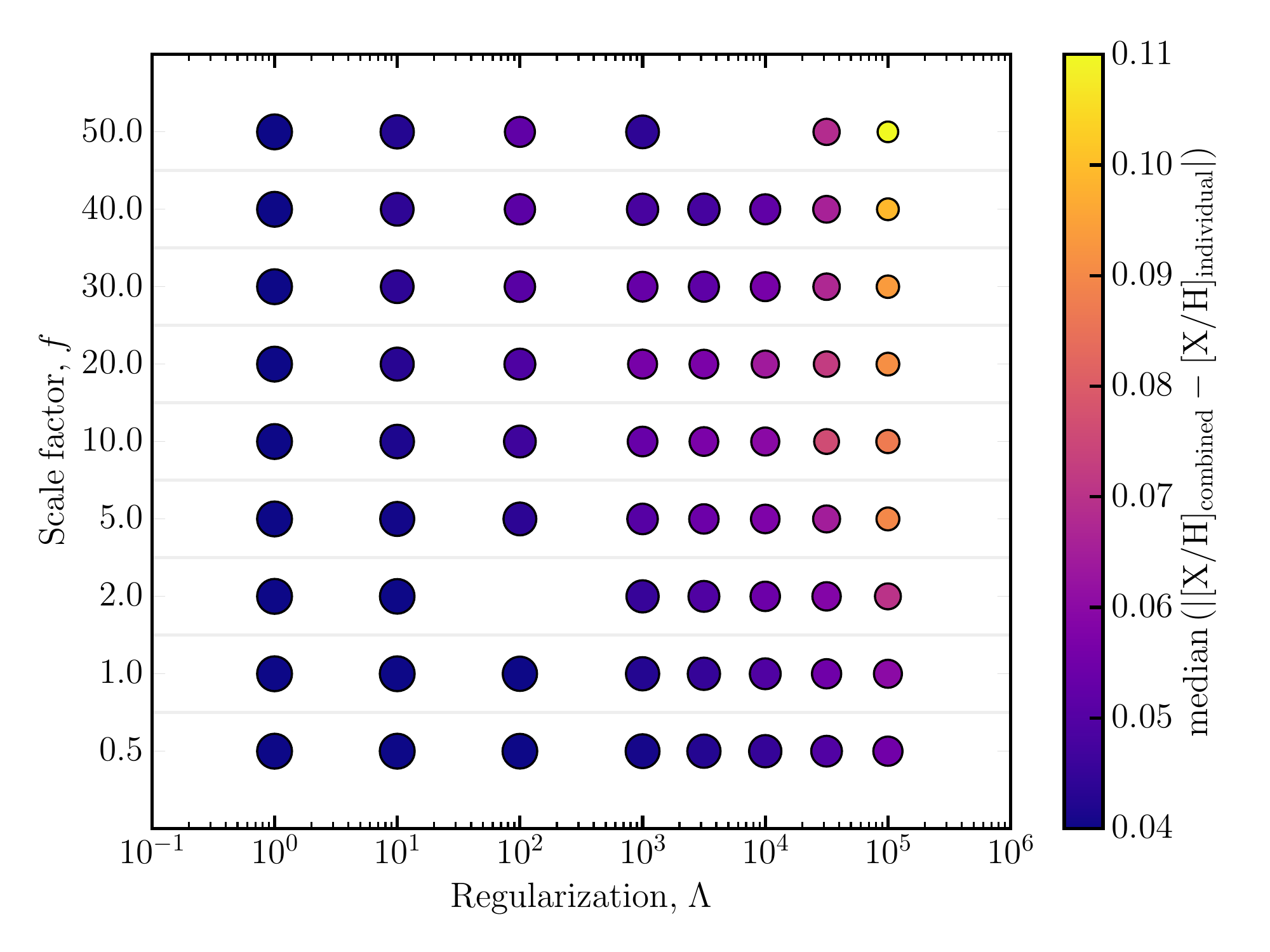}
\caption{The median absolute abundance difference (over all abundance labels)
between that inferred from individual observations in the validation set, and
that inferred from the high S/N combined spectrum.
\label{fig:gridsearch-mad-all-elements}}
\end{figure}

While sparser models are preferred, our ultimate goal is to have an 
interpretable model that predicts spectral fluxes and returns precise stellar
labels.  In Figure \ref{fig:gridsearch-mad-all-elements} we show the median
absolute deviation in abundance labels, measured between individual and combined
spectra for stars in the validation set.  This is an internally-consistent
check for label recovery at low S/N: we will validate our high- (and low-) S/N
label determination against \aspcap\ values in Section \ref{sec:model-validation}.  It is
clear that a combination of decreasing $f$ with increasing $\Lambda$ recovers
labels for validation set stars with high precision.  At $\Lambda \approx 10^3$,
varying $f$ between 0.5 and 50.0 results in a marginal change in the validation
set precision (0.04~dex to 0.06~dex).  Figure \ref{fig:gridsearch-mad-all-elements}
suggests that at $\Lambda = 10^3$, $f = 0.5$ has comparable performance in
label recovery as a model with the same $\Lambda$ and $f = 50.0$.  While this
behaviour was observed in the recovery of some abundance labels, it was not seen in all. 
The behaviour differed for each label.  However, lower scale factors were favoured by all labels.

\begin{figure}[p]
\centering
\includegraphics[width=\textwidth]{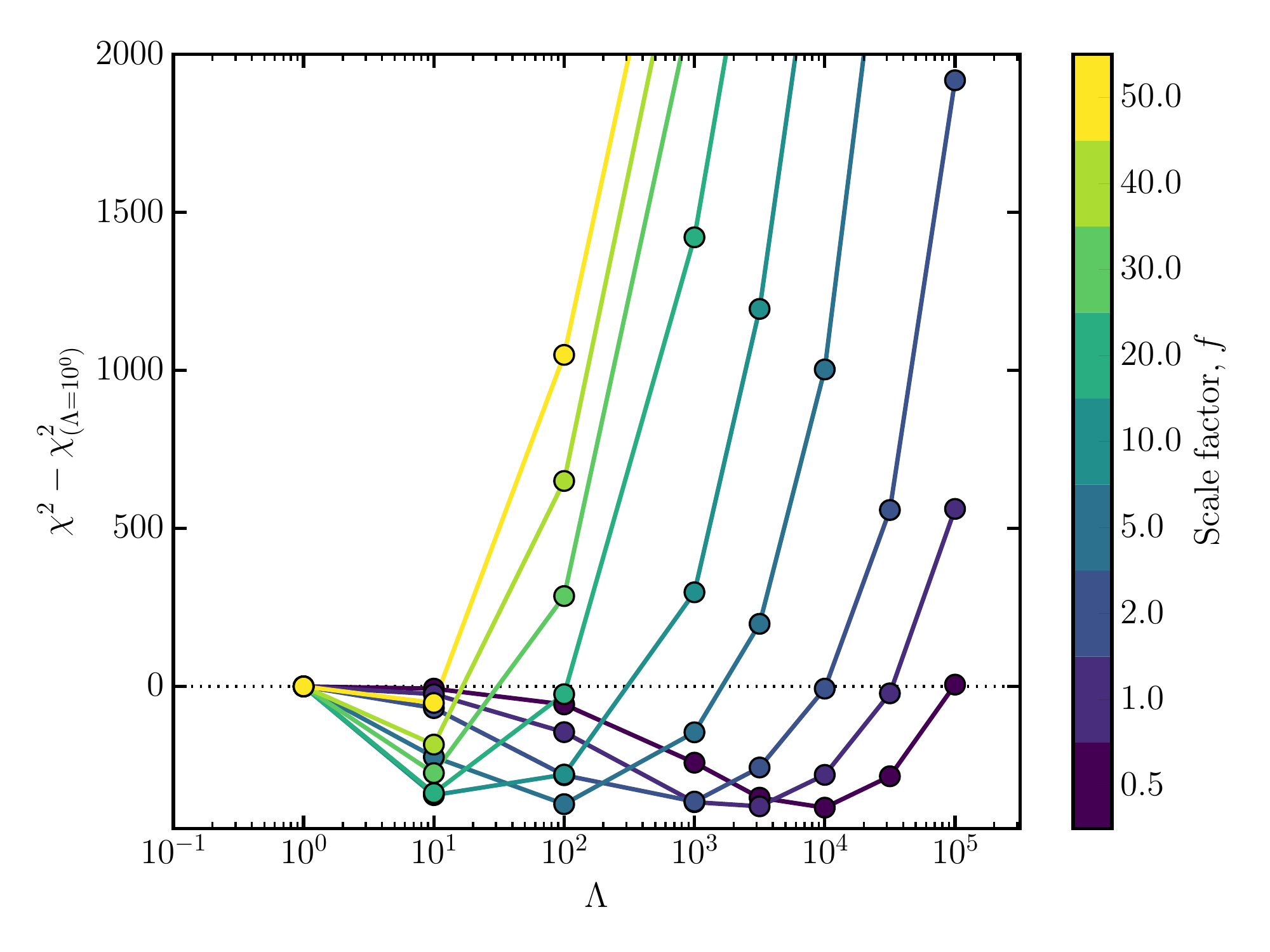}
\caption{The total $\chi^2$ (i.e., the \emph{test scalar}) relative to the $\Lambda = 10^0$ model for all validation set stars for different combinations of the hyper-parameters $\Lambda$ and $f$.
 \label{fig:gridsearch-test-scalar}}
\end{figure}

\begin{figure}[p]
\centering
\includegraphics[width=\textwidth]{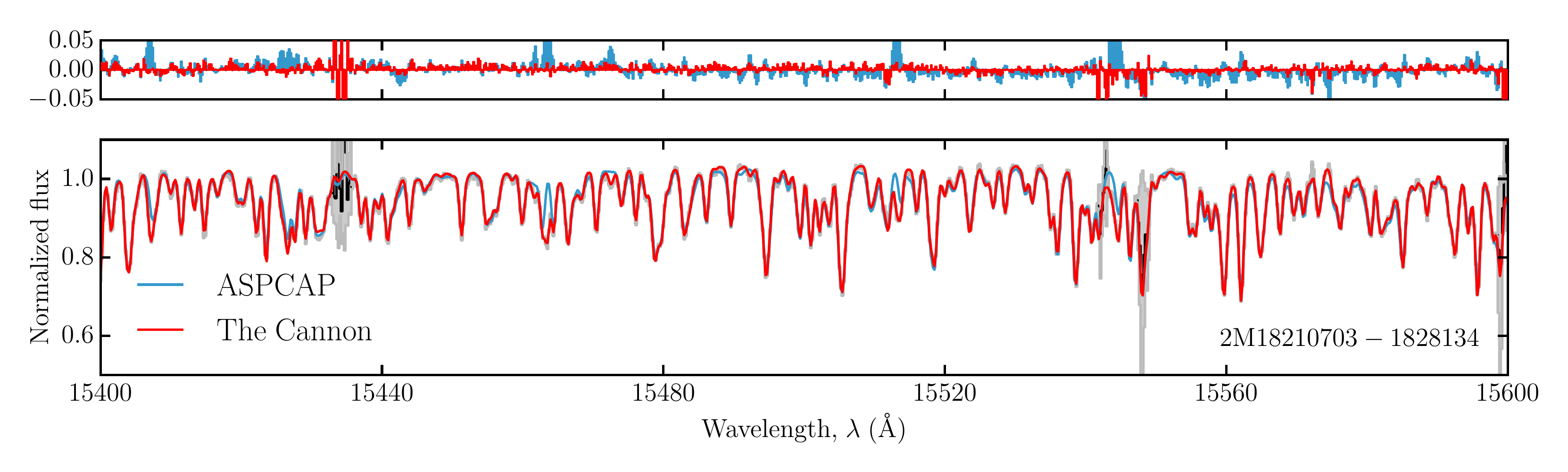}
\caption{A portion of the (\emph{Cannon}-)normalized spectrum for 2M18210703-1828134,
a randomly selected star in the validation set.  The best-fit model spectra
for \aspcap\ and \TheCannon\ are shown. The residuals are plotted in the top panel.\label{fig:correctness}}
\end{figure}

As a final heuristic to guide our choice of $\Lambda$ and $f$, we examined the
performance in predicting spectral fluxes for all validation set spectra.  Here 
we predicted stellar fluxes for all validation set stars (using the \aspcap\ 
labels) and calculated the total $\chi^2$ difference.  The results are shown in
Figure \ref{fig:gridsearch-test-scalar} compared to the $\chi^2$ for the least-regularized case ($\Lambda = 10^0$).  At increasing regularization strength
the models demonstrate better predictive power in spectral fluxes, reflected by
a lower total $\chi^2$ value.  As expected, the $\chi^2$ minimum is dependent on
the scale factor.  While regularized models with lower $\chi^2$ values clearly
predict spectral fluxes more accurately (a comparison with \aspcap\ predicted
fluxes is shown in Figure \ref{fig:correctness}), this heuristic only gives a 
weak limit on our choice of hyper-parameters: \emph{any} regularized model with
a lower $\chi^2$ value than seen in the unregularized case is objectively a 
better model because it predicts stellar spectra more accurately.

Our grid search has revealed that a regularization factor of at least $\Lambda = 10^3$ is
required to produce a sufficiently sparse model.  Generally we find that low
scale factors yield marginally better behaviour in recovering abundance labels
at low S/N.  At $\Lambda = 10^3$, the precision in abundance labels would 
suggest that a scale factor anywhere between $f = 0.5-5.0$ is reasonable. 
However the total $\chi^2$ for validation set fluxes shows $f = 2.0$ to be a far
better model (in terms of predictive power in spectral fluxes) than $f = 0.5$
or $f = 5.0$.  Thus, on the basis of these metrics, we adopt $\Lambda = 10^3$
and $f = 2.0$ as our \emph{regularized model} hyper-parameters for the
remainder of this work.

\subsection{Validating the regularized model}
\label{sec:model-validation}

\begin{figure}[p]
\centering
\includegraphics[width=0.8\textwidth]{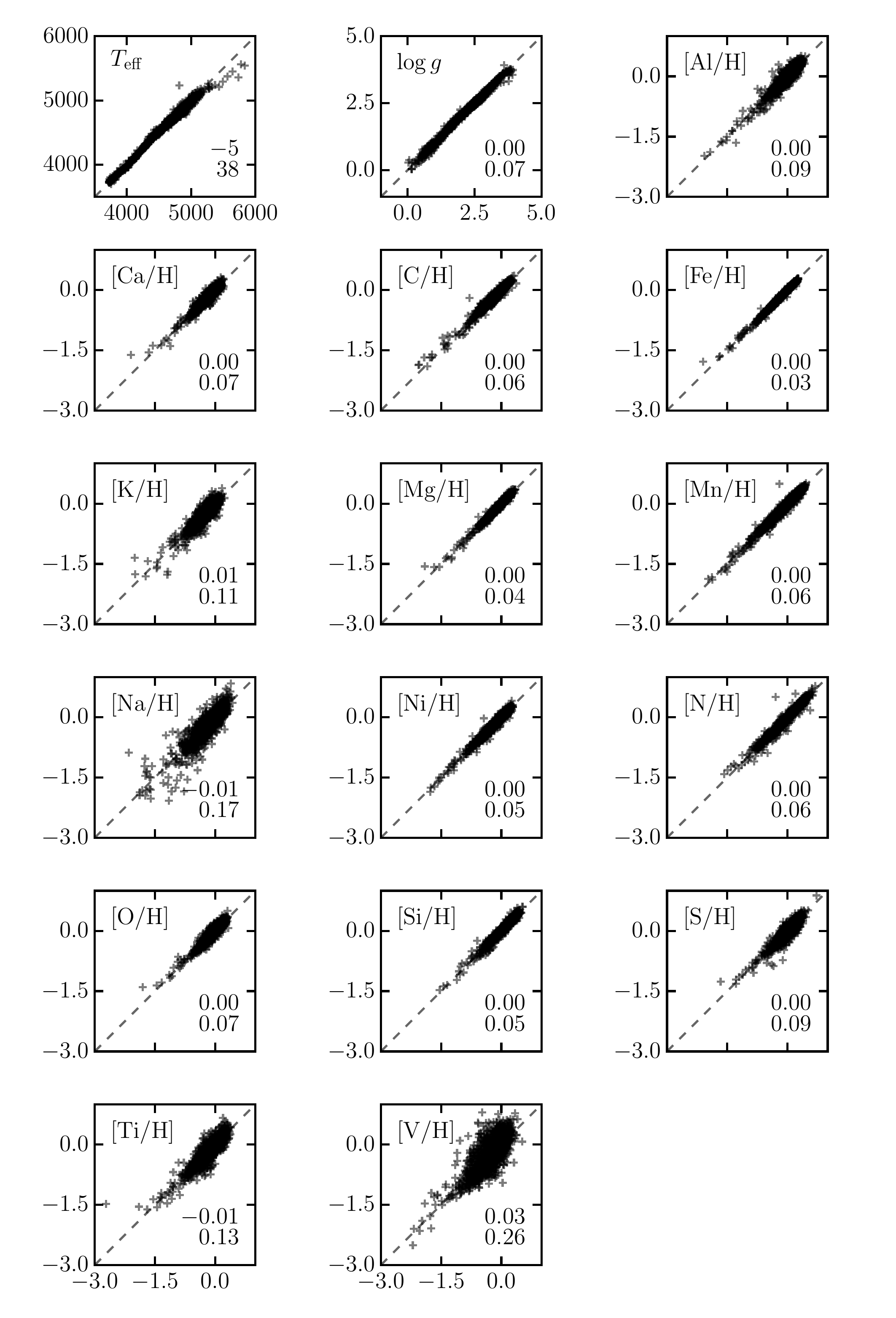}
\caption{Comparisons between \aspcap\ (x-axis) and \TheCannon\ (y-axis)
labels for the 1,460 high S/N stars in the validation set.  For clarity
purposes the label name is shown within each panel, as is the mean
and standard deviation of the residuals.  A
dashed one-to-one line is shown.
\label{fig:regularized-model-validation}}
\end{figure}

We re-trained the 17-label model (without requiring $s^2 = 0$) using our adopted
hyper-parameters.  In this and subsequent sections we validate our model through
various experiments.  Our first test is to examine how well we can recover
labels for the validation set stars.  Recall that these stars are a random 10\%
subset of the reference set and were not used for training.  Comparisons between
the \aspcap\ validation set labels and our own are shown in Figure
\ref{fig:regularized-model-validation}, where we find that the model shows good
agreement with the \aspcap\ labels.  No strong biases are seen, and the standard
deviation of the residuals is low for most abundance labels.  The largest
disagreement is seen in [V/H] and [Na/H], in different regimes of label space.
While [Na/H] starts to become discrepant at [Na/H] $< -1$ (perhaps because we
are reaching the noise floor for this label), [V/H] becomes increasingly discrepant
near [V/H] $\approx 0$.

The [V/H] abundance labels are problematic in the \aspcap\ pipeline 
because the signature is subtle even at high S/N ratios: there are very few
spectral lines, most of which are blended with stronger lines.  Indeed, we noted
very obvious issues in the [V/H] labels in the reference set which we tried to
account for (e.g., Section \ref{sec:training set}).  Although our [V/H]
label precision is worse than other abundance labels, it is limited by the
training set.

We report formal errors in all labels provided by \TheCannon. At test time the
Jacobian matrix of partial derivatives is estimated at the optimized solution,
which we multiply by the residual variance to obtain a formal covariance matrix.
Errors in each label are taken as the square-root of the diagonal entries along
the covariance matrix. We stress that the errors listed in Table \ref{tab:the-good-stuff}
are formal errors only: they do not encapsulate all aspects of the error budget
that are known (or unknown) to us. An uncertainty floor in each label is estimated
in the following Section, which ought to be added in quadrature with the formal
errors.

\subsection{Label recovery as a function of signal-to-noise}
\label{sec:label-recovery-snr}

Here we seek to understand the capability of our 
regularized model to recover labels as a function of S/N.  The validation set includes stacked spectra for each star
 as well as the individual normalized observations.  We measured labels for every spectrum.  This permits us
to understand how well we can recover labels (to some precision) for a given 
S/N.  Crucially, this experiment provides a very model-independent metric for
performance.  At low S/N ratios the label errors will be dominated by random
uncertainties.  In the high S/N regime, systematic uncertainties are more 
relevant.  Knowing exactly where this ``turnover point'' occurs from systematic
to random uncertainties is a useful metric to compare analysis pipelines.

\begin{figure}[p]
\centering
\includegraphics[width=0.8\textwidth]{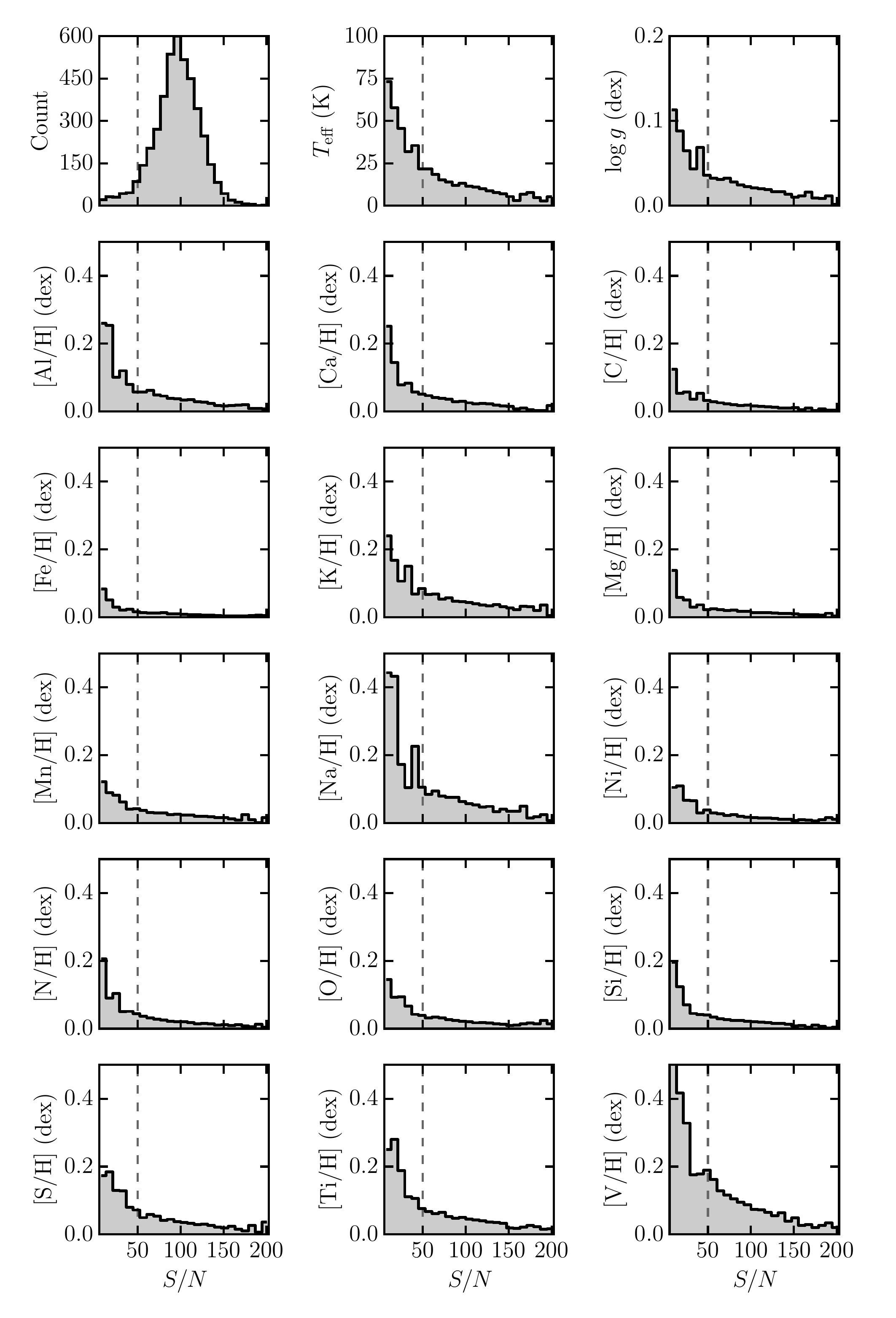}
\caption{\emph{Top left panel}: The distribution of S/N values for spectra of individual observations in the validation set. \emph{All other panels}: The median absolute difference between the measured label for an individual observation, and the measured label from the high S/N combined spectrum for the same star.  All panel have common bins, and all abundance labels have common axis limits.  The dashed line indicates the minimum S/N for any \emph{stacked} \apogee\ spectrum in Table \ref{tab:the-good-stuff}, implying the precision floor achievable for all \apogee\ stars.\label{fig:label-recovery-snr}}
\end{figure}

The results of our label recovery experiment from \emph{individual visit} spectra of
validation set stars are shown in Figure \ref{fig:label-recovery-snr}.  
At $S/N \gtrsim 50$ the discrepancies in all
labels are either flat or taper towards zero, indicating that we are dominated
by systematics.  This is an important transition point, because \emph{stacked} spectra for all bonafide giant
stars (Section \ref{sec:results}) have $S/N \geq 50$.  In this regime we 
recover labels from within 0.01~dex ([Fe/H]) to 0.16~dex ([V/H]) for all 
abundance labels.  The worst performance is again found for [V/H], by a factor of
two: the second-worst label at $S/N \geq 50$ is [Na/H] with 0.08~dex.  These
values are listed for all labels in Table \ref{tab:error-floors}, where we advocate their use as a representative error floor that should be added in quadrature with the formal errors. We stress that this is not a thorough accounting of the error budget.

It is clear that the regularized model recovers labels with good precision even
in the presence of substantial noise.  In our experience the recovered 
precision presented here is higher than most physics-driven analysis pipelines.  
The reason for this is not completely understood (by us), because both approaches 
(optimizing a data-driven or physics-driven model) rely on least-squares 
fitting, which will usually follow the $1/\sqrt{S/N}$ behaviour we see in Figure 
\ref{fig:label-recovery-snr} unless there are strong non-linearities in the model.
Unfortunately we cannot show the same comparison for \aspcap\ labels as a 
function of S/N in Figure \ref{fig:label-recovery-snr} as \aspcap\ labels are not
publicly available for individual visit spectra.

\section{Results}
\label{sec:results}

Our experiments have demonstrated that a data-driven model for stellar spectra
can be reliably extended to high dimensionality in label space.  We have further
shown that the regularization hyper-parameters can be simplified to just two
hyper-parameters that can be set heuristically.  This yields a sparse, interpretable 
model that recovers labels with high precision at low S/N.  While this is 
reassuring, it is less relevant for our results as all stacked \apogee\
spectra for bonafide giants (see below) have $S/N \gtrsim 50$, well into the 
regime where we are advantageously dominated by systematic uncertainties.

We have used our regularized model to measure (test) labels of 150,677 \apogee\ 
spectra, all normalized and stacked using the method in Section 
\ref{sec:training set}.  In addition to the model being effective,
the test step is very fast: our pure-\texttt{Python} implementation returned 
\emph{17 labels for all 150,677 spectra in just 28 minutes} of wall-time 
on a small research cluster in Cambridge.  These were free 
and otherwise unused resources; no dedicated computing assets were required.  
This pace is also projected to increase, as the test step did not include 
analytic derivatives $\delta\vectheta/\delta{}y_j$, which are now implemented
in our open-source code for any polynomial vectorizer.  The test-step 
optimization is not convex because the vectorizer contains
quadratic label terms.  For this reason we ran the optimization from nine
different initialization points, chosen to sparsely cover the range of
$\Teff$, $\logg$, and abundance labels in the training set.  Of the nine
optimizations, we adopted the end result with the lowest $\chi^2$ value.

\begin{figure}[p]
\centering
\includegraphics[width=0.8\textwidth]{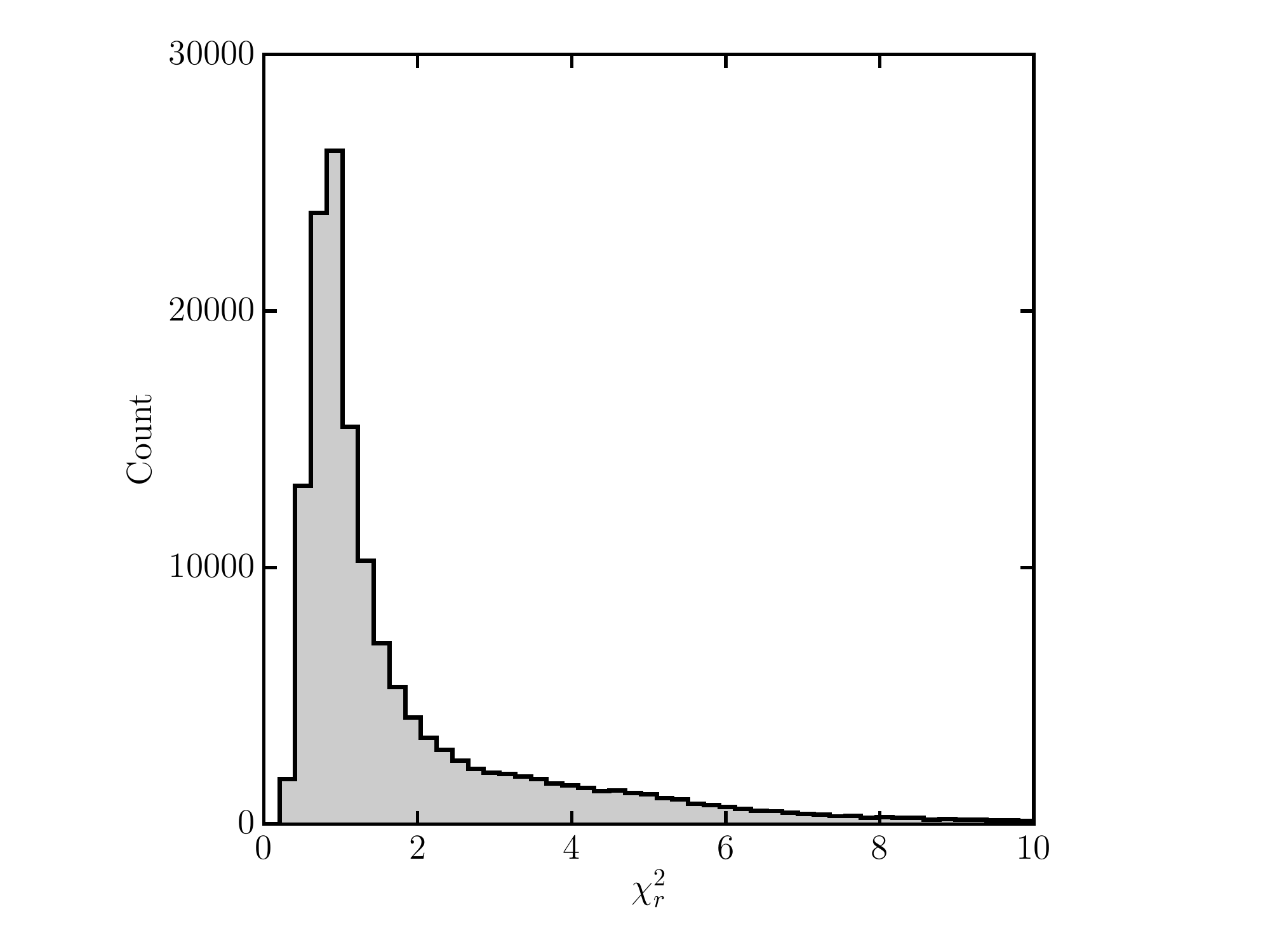}
\caption{Reduced $\chi^2$ values for all stars (stacked spectra) in \apogee\ \dr.\label{fig:chisq-test set}}
\end{figure}

\begin{figure}[p]
\includegraphics[width=\textwidth]{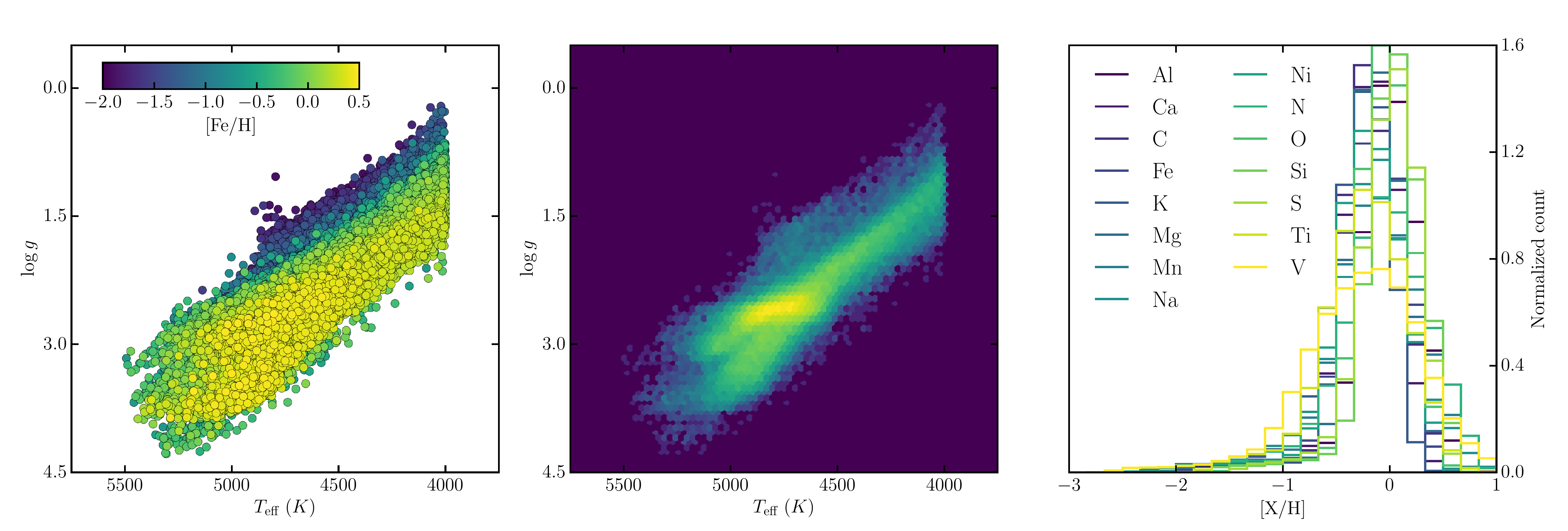}
\caption{A temperature-surface gravity diagram for all 87,563 stars in the test set that pass our quality criteria (left). The centre panel shows a logarithmic density plot for the same stars. Abundance label distributions are shown in the right-hand panel.\label{fig:test set-hrd}}
\end{figure}

The training set only includes giant stars, but the \apogee\ \dr\ includes 
giants and dwarfs.  Therefore we exclude results with $\chi_r^2 > 3$, stars without a $\Teff$ label from \aspcap, or stars with \aspcap\ labels outside of the bounds $5500 > \Teff > 4000$ or $\logg < 3.9$.
The distilled sample contains 87,563 giant stars, where we report $\Teff$,
$\logg$, and 15 abundance labels.  The distribution of $\chi^2$ values for
all 150,677 combined spectra are show in Figure \ref{fig:chisq-test set}.  The
labels in the distilled sample follow expectations from stellar astrophysics,
and include stars that are marginally outside the training set.  For example,
the \aspcap\ labels include a strict cut in $\Teff$ at 3600~K, but we reliably
recover labels beyond this boundary.  Figure \ref{fig:test set-hrd} presents a
few different label projections for the distilled sample, indicative of the
boundaries and distribution of our labels.  The full complement of 17-labels
for 87,563 stars is provided in Table \ref{tab:the-good-stuff}, in addition
to quality metrics.

\begin{figure}[p]
\centering
\includegraphics[width=0.8\textwidth]{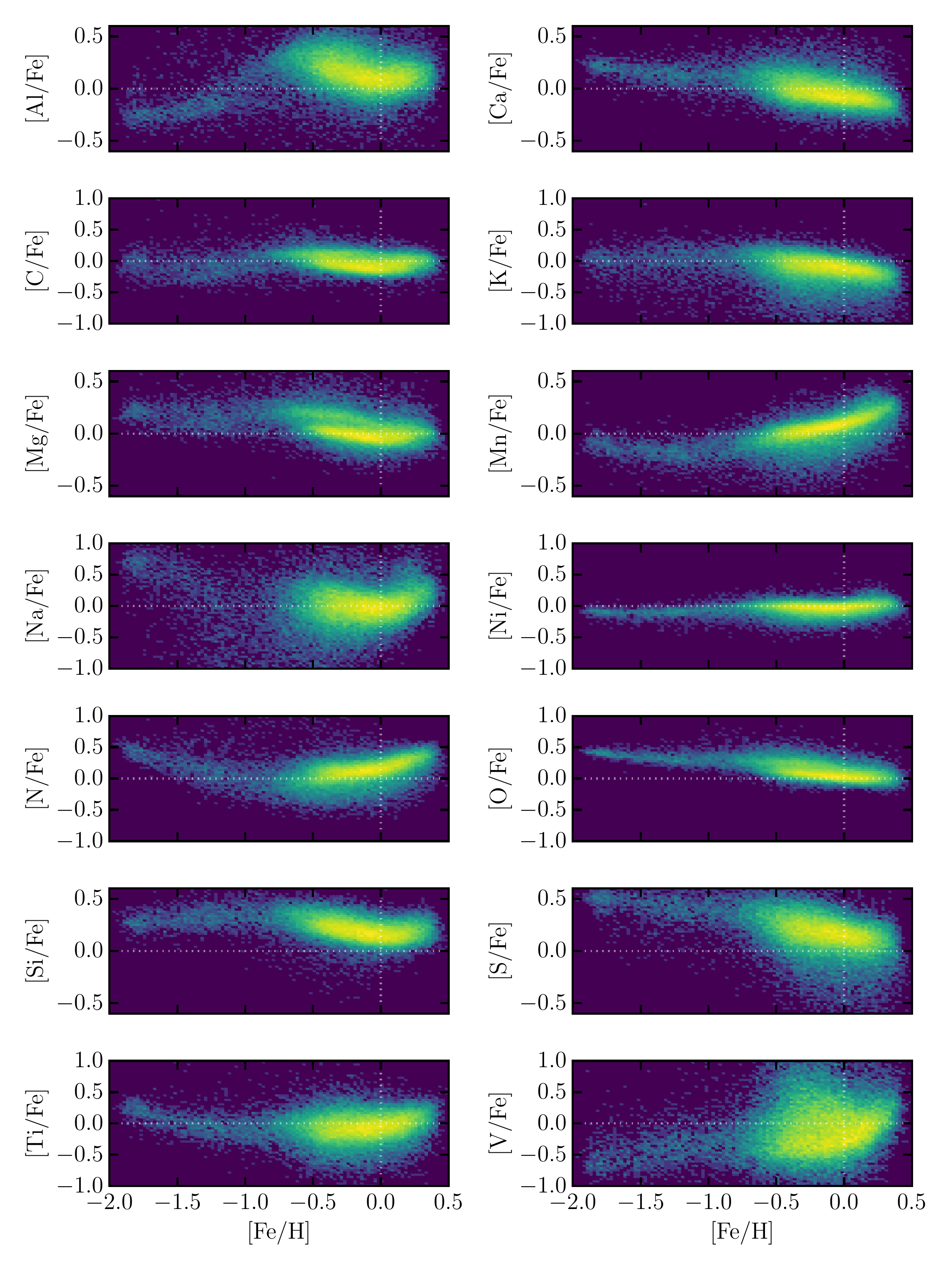}
\caption{Logarithmic density for all abundance labels with respect to iron ([X/Fe]), revealing the Galactic  enrichment with respect to Fe for all elements. Dotted lines indicate Solar abundances.\label{fig:gce}}
\end{figure}

Our abundance labels are consistent with detailed studies
of galactic chemical evolution.  In Figure \ref{fig:gce} we show our abundance
labels (with respect to iron) of $\alpha$-capture (e.g., O, Mg, Ca, Si), odd-Z
(Al, Na), light (C, N), and Fe-peak (Ti, Mn, Ni) elements for all bonafide
giant stars in \apogee\ \dr.  These abundances trace nucleosynthetic pathways
from different sources (e.g., supernovae, AGB stars) and reflect the environmental
conditions at the time of their formation.  In Section \ref{sec:discussion} 
we will briefly discuss the implications and interpretations of these abundance
labels.

\begin{figure}[p]
\centering
\includegraphics[width=0.7\textwidth]{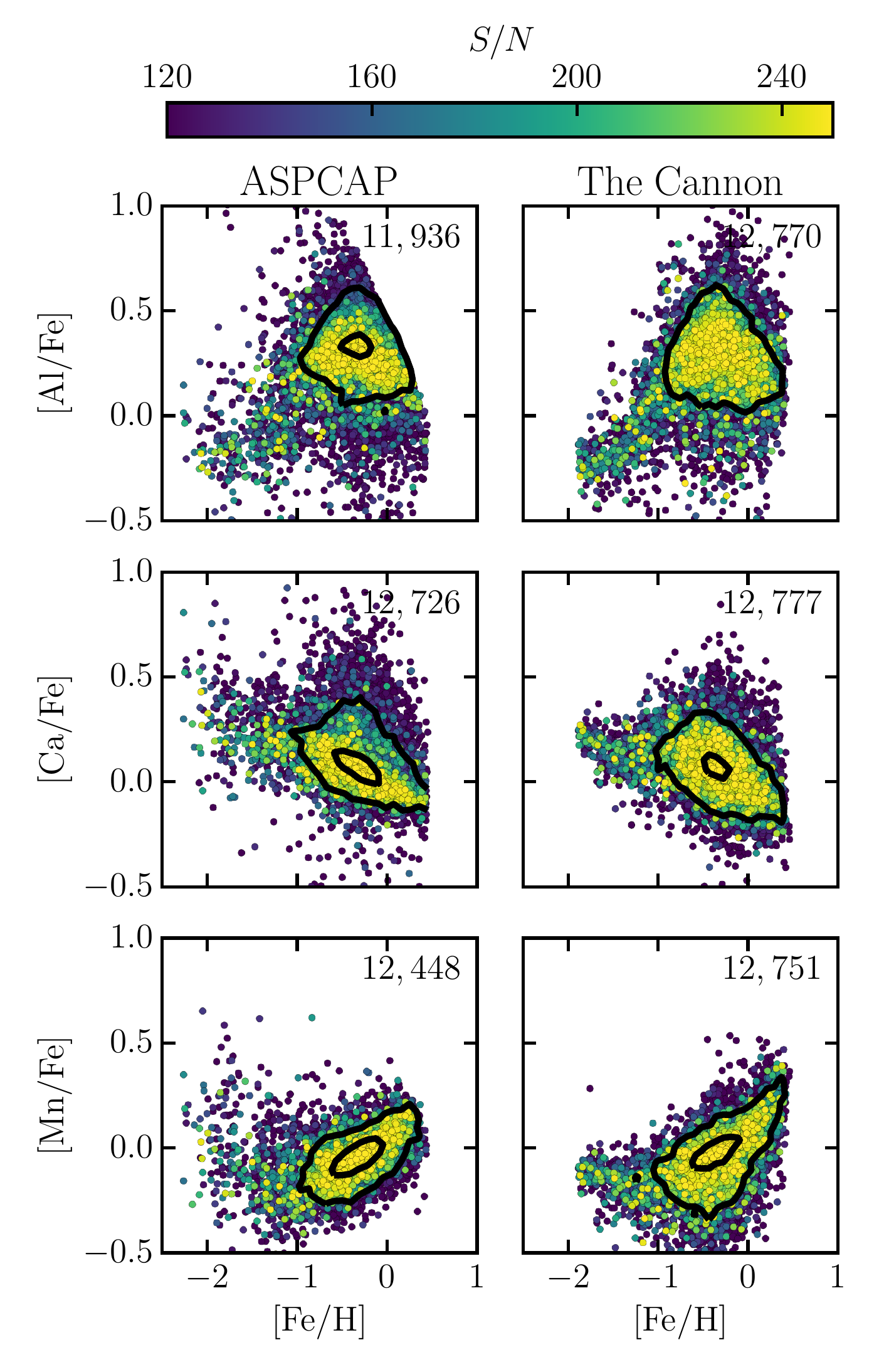}
\caption{Abundance labels of three element groups (odd-Z, $\alpha$, and Fe-peak) for stars in a high [$\alpha$/Fe] abundance sequence. Labels from \aspcap\ (left) and \TheCannon\ (right) are shown for the same set of stars. Stars are colored by their S/N ratio to highlight that the spread in the \aspcap\ label distribution is strongly impacted by stars with modest S/N ($<120$). The same result is not seen with \TheCannon.\label{fig:high-alpha-sequence}}
\end{figure}

We trained \TheCannon\ using high-fidelity labels from \aspcap.  As expected,
our labels agree excellently with \aspcap\ for stars with high S/N ratios.
This is not the case for stars with S/N ratios below 200.  Here we present
abundance label projections for some astrophysically interesting subsets of
the \apogee\ sample, and show comparisons between \TheCannon\ and \aspcap.
Our first comparison is shown in Figure \ref{fig:high-alpha-sequence}, where 
we have selected a sequence of stars with
high [$\alpha$/Fe] abundance ratios and shown three well-behaved (precise)
abundance labels that trace different nucleosynthetic pathways.  Points are
coloured by their S/N ratio, and the black lines indicate density contours.

The differences between \TheCannon\ and \aspcap\ projections in Figure \ref{fig:high-alpha-sequence} are striking.  Labels
from \aspcap\ have a much larger range than those from \TheCannon,
however nearly all stars that fill this difference in label range are those
with low S/N ratios.  This implies that either those stars with lower S/N
are sampling a different (presumably more distant) part of the Milky Way,
or the precision in \aspcap\ labels degrades much worse at low S/N ratios than \TheCannon.
Our validation tests have demonstrated that our label precision remains
approximately constant (systematic-dominated) at $S/N \gtrsim 50$ (i.e.,
for all combined spectra of bonafide giants in \apogee\ \dr).  For these reasons we argue
that the difference in abundance projections between \TheCannon\ and
\aspcap\ for low S/N stars in the [$\alpha$/Fe] set are likely because \aspcap\
yields imprecise results for spectra with S/N below $\sim$200.  Although
the imprecise \aspcap\ results can presumably be discarded with some 
quality cuts, a substantial fraction of the \apogee\ sample would have
to be removed.

Globular clusters are excellent laboratories for us to validate
our abundance precision and label interpretability.  Indeed, it is our
firm belief that there is sufficient new information \emph{just} in our
globular cluster abundance labels that would warrant its own publication.
We encourage others to pursue this; here we simply comment on how our results 
compare to \aspcap, and in the next Section we will discuss how our
results compare to existing studies on globular clusters, and 
expectations from stellar evolution.  Here we discuss three globular
clusters that have a considerable number of bonafide members in \dr:
M~15, M~53, and M~92.

\begin{figure}[p]
\centering
\includegraphics[width=0.35\textwidth]{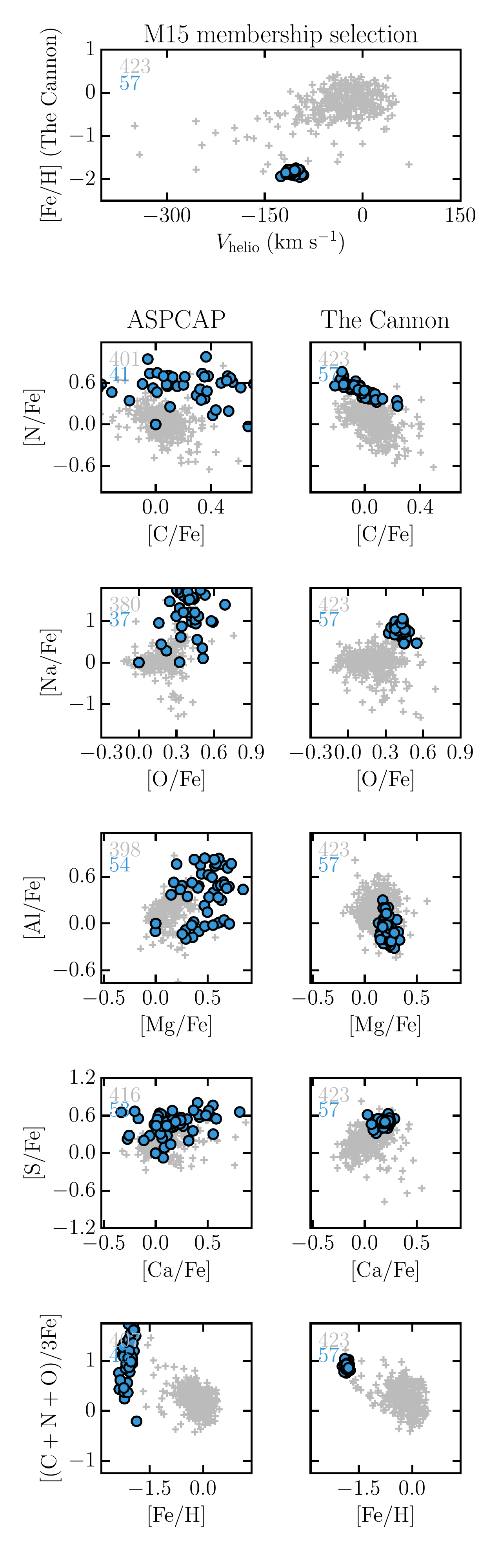}
\caption{Abundance labels from \aspcap\ and \TheCannon\ for candidate
members of the metal-poor globular cluster M~15 (NGC~7078).  The top panel
shows heliocentric velocity and metallicity, which we use to separate
bonafide cluster members (blue) from field stars (grey). Panel limits
are set by the dynamic range of \TheCannon\ labels.  The number of stars
within those limits (from \aspcap\ and \TheCannon) are shown.\label{fig:m15-comparison}}
\end{figure}

\begin{figure}[p]
\centering
\includegraphics[width=0.35\textwidth]{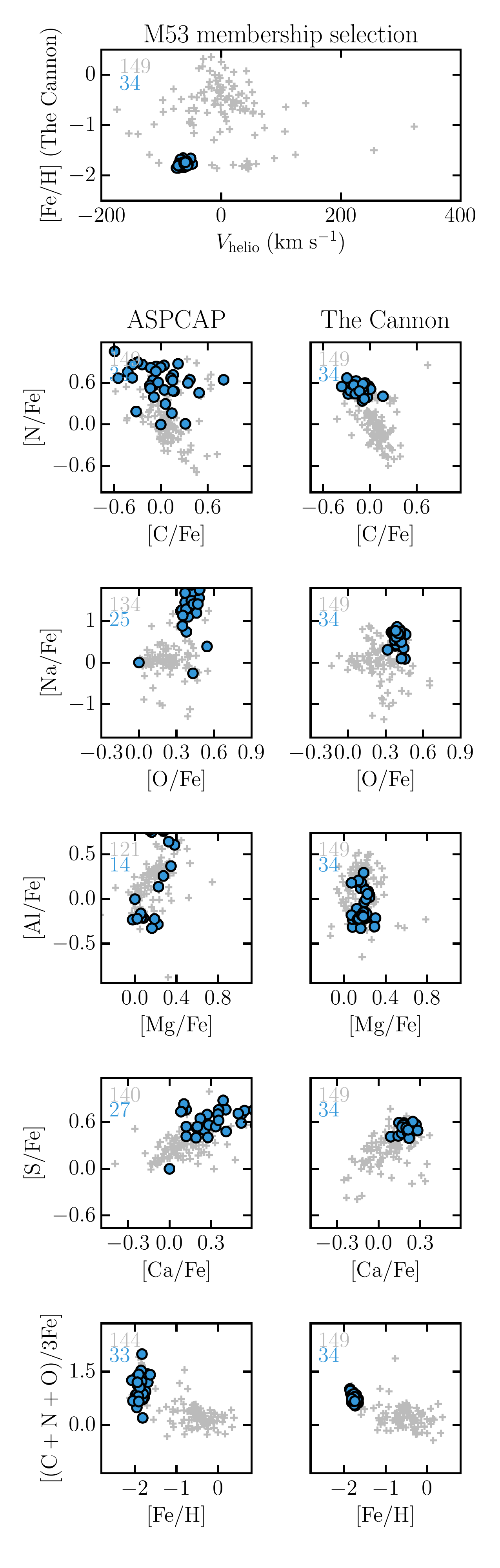}
\caption{Label comparison between \aspcap\ and \TheCannon\ for M~53.
Markings are the same as per Figure \ref{fig:m15-comparison}.
\label{fig:m53-comparison}}
\end{figure}

\begin{figure}[p]
\centering
\includegraphics[width=0.35\textwidth]{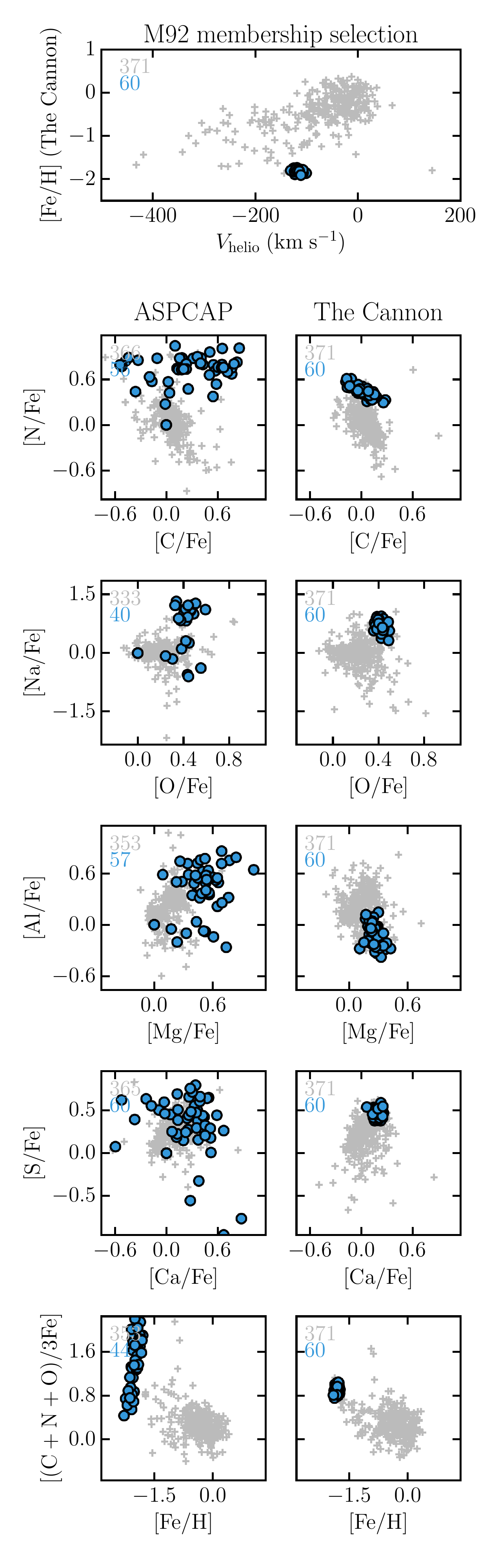}
\caption{Label comparison between \aspcap\ and \TheCannon\ for M~92.
Markings are the same as per Figure \ref{fig:m15-comparison}.
\label{fig:m92-comparison}}
\end{figure}

In each cluster we attributed stars to be bonafide members based on the
\aspcap\ radial velocity and \TheCannon\ [Fe/H] label.  The top panel of
Figures \ref{fig:m15-comparison}-\ref{fig:m92-comparison} shows our 
membership selection with respect to all other \apogee\ giants in the
same field (i.e., those that share the same `\texttt{FIELD}' entry).
Although our membership criteria may suffer from mild contamination of
field stars, this does not significantly impact our comparisons to 
\aspcap\ labels or any discussion in the following Section.  The 
remaining panels of Figures \ref{fig:m15-comparison}-\ref{fig:m92-comparison} show individual abundance projections
using \aspcap\ labels (left) and those from \TheCannon\ (right). 
In any projection, the abundance labels from \TheCannon\ show a smaller
spread than \aspcap\ for the same stars.  In most cases the mean cluster
labels between \aspcap\ and \TheCannon\ are similar, with correlations
in the same direction. The only difference is that smaller label
distributions are found by \TheCannon.  The most striking comparison 
is seen for M~15 (Figure \ref{fig:m15-comparison}).

\section{Discussion}
\label{sec:discussion}

We have demonstrated that a data-driven model for stellar spectra can be
reliably extended to high dimensionality in label space. We return high
precision labels for \apogee\ spectra across the giant branch, including
those with low S/N ratios. For these noisy data, the abundance labels returned
by our model for low S/N ratios show similar behavour to what is seen in the high
S/N ($>250$) sample.  However our abundance labels from noisy data are
significantly different from \aspcap. In this noisy regime
there are good reasons to believe \TheCannon\ abundance labels are more
reliable.  This has a prominent impact on abundance labels for globular
cluster stars, as our results suggest much narrower intrinsic abundance
spreads ($\sim$3-4 times smaller).

There are (at least) two possibilities to explain the discrepancies in
cluster abundances between \aspcap\ and \TheCannon.  The first is that
there is a large underlying intrinsic abundance spread in these elements
for each cluster, and -- for whatever reason -- \TheCannon\ is returning
systematically inaccurate abundances.  The alternative is that the
intrinsic cluster abundance spreads \emph{are} smaller than what \aspcap\
labels suggest, and the larger cluster spreads in \aspcap\ abundances are
simply due to large uncertainties in individual abundances.  On the basis
of our validation tests, we argue the latter to be a more plausible
explanation for these abundance discrepancies.

The overall metallicities ([Fe/H] label) we find for the globular
clusters examined are in satisfactory agreement with the literature.  In
general our [Fe/H] label is more metal-rich than what is quoted in existing cluster
compilations \citep{Harris_1996}.  The difference seems to be an imprint from
the \aspcap\ labels at low-metallicity, as the \aspcap\ labels for 
stars we assign as cluster members are also more metal-rich than
literature sources. This deviation may itself be linear in the 
\aspcap\ [Fe/H] scale, as the agreement is very good for 
metal-rich clusters in our sample.  For example, for M~3 we find a cluster mean and standard
deviation of $[{\rm Fe/H}] = -1.38 \pm 0.09$, just 0.12~dex more metal-rich
than the literature mean \citep{Harris_1996}.  Similarly we find M~13 to have
[Fe/H] $= -1.45 \pm 0.09$ whereas the same source quotes [Fe/H] $= -1.53$.
However at the metal-poor end the disparity becomes severe, where we find
a 0.5~dex offset with the literature for M~92 ($[{\rm Fe/H}] = -1.81 \pm 0.04$).

While these mismatches are important to examine, our tests suggest they
are a reflection of two compounding issues: a relative paucity of metal-poor
stars in the training set, and a suggestion of a linear-in-[Fe/H] systematic
trend in \aspcap\ [Fe/H] labels.  A more thorough comparison of metal-poor
stars in \aspcap\ may help resolve these issues.  Nevertheless, while
\TheCannon\ abundance scale may be slightly inaccurate (offset from the
literature consensus) for metal-poor stars, this issue does not impact
our precision in abundance labels that we report for these clusters.

The anti-correlation we see between [C/Fe] and [N/Fe] abundances in M~15 is
expected from the CN-cycling of stellar material.  The labels (and correlation
strength) are in excellent agreement with high-resolution studies on this cluster
\citep{Cohen_2005}.  Indeed, while we find an intrinsic spread in these abundance
labels, we find that the spread is much smaller than considered previously.  We also
corroborate works that find no evidence of bi-modality in C and N 
abundance labels, a signature commonly found in metal-rich globular clusters.
While the \aspcap\ labels show a substantial spread in total C+N+O abundances
for M~15, our results imply that CNO-cycling has maintained an approximately
constant sum abundance of these elements, consistent with recent works
\citep{Meszaros_2015}.

The reference set described in Section \ref{sec:training set} was not
purposefully constructed to include stars in \emph{any} globular
cluster.  We selected stars that met strict quality criteria, and
discarded stars with questionable label fidelity.  Given the
abundance (anti-)correlations we confirm from other high-resolution
studies, our construction of the reference set has a number of
implications.  First, \emph{the anti-correlations seen in globular 
clusters here and elsewhere are not dominant in our training set}.
Indeed, the lack of globular cluster (metal-poor) stars in our 
training set partially explains the [Fe/H] discrepancy we see with
respect to the literature (i.e., very few metal-poor stars in the
training set).  However as a consequence the regularized model is 
successfully measuring sets of labels (e.g., chemical fingerprints) that are 
\emph{very different} to what the model was actually trained on.  
Even though these chemical signatures are distinct from the training
set, an experimental astrophysicist expects to see them based on previous works.  For this
reason it is very pleasing to see these signatures because (amongst
other things) it reflects the interpretable nature of our model: the spectral
derivatives \emph{do} have physical meaning, allowing for the model
to recover label patterns of high astrophysical interest that are 
substantially different from the training set.

Our measured labels have good astrophysical
interpretations that are consistent with expectations from stellar
evolution.  The photospheric abundances of [C/H], [N/H] and [O/Fe] 
are expected to change as a giant star experiences CNO cycling.  This pattern is
present in our data, and to a lesser extent it is visible in the
\aspcap\ labels.  Whilst these abundances vary, the sum of these
elemental abundances is expected to remain roughly constant for a
giant star, regardless of its evolutionary state.  This is not 
seen in the \aspcap\ labels in any globular cluster described here.  
A huge spread in total [(C,N,O)/3Fe] abundance labels is seen, 
which is perhaps reflective of the sum of large uncertainties in
each \aspcap\ C, N, and O abundance label.  In contrast, \TheCannon\
labels remain roughly constant for each globular cluster.

\begin{figure}[p]
\includegraphics[width=\textwidth]{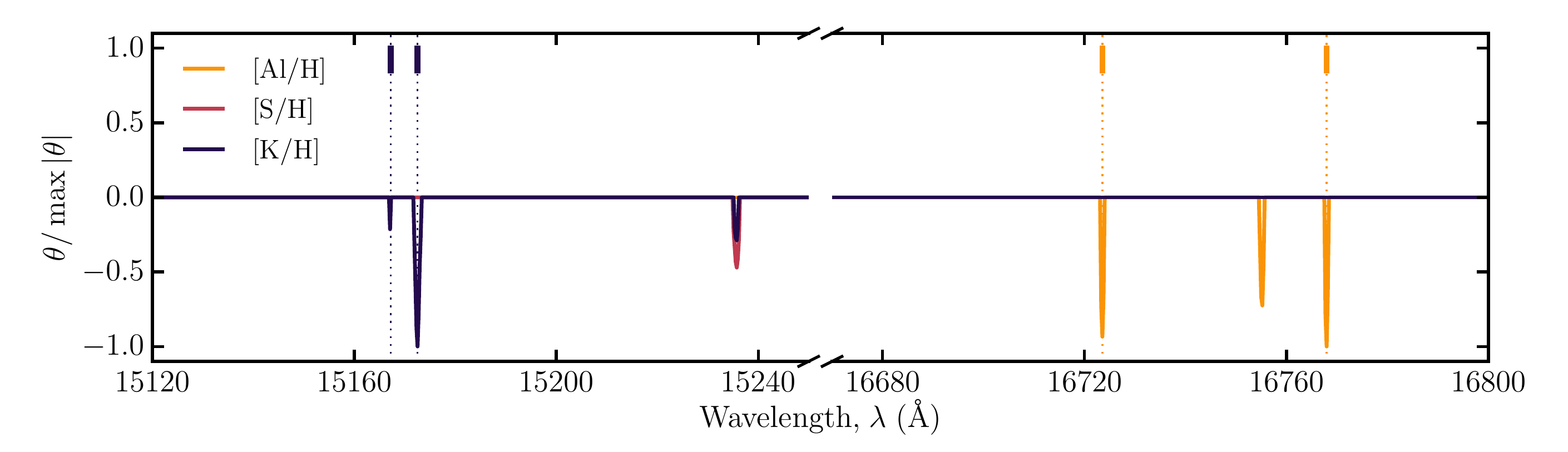}
\caption{The normalized (to the maximum derivative value at any $\theta$) first-order spectral derivatives for [Al/H], [S/H], and [K/H] from our regularized \emph{Cannon} model. Vertical markings (and their colors) correspond to strong spectral lines used by \citet{Smith_2013}. For the sake of clarity we only show $\theta$ values above a background threshold.\label{fig:line-identification}}
\end{figure}

In addition to returning precise abundance labels, the regularized model
we have presented is interpretable.  Machine learning techniques can be extremely
useful, but they are frequently limited in their application because they
usually do not have internal components that are interpretable.  This is
not the case for \TheCannon: the spectral coefficients $\vectheta_0$
represent the baseline spectrum, and the first-order coefficients 
($\vectheta_{1-17}$ in this model) represent the spectral derivatives with
respect to each label.  They indicate where specific labels (or elements)
contribute to the spectrum.  In Figure \ref{fig:line-identification} we
demonstrate an example of this, where the first-order derivatives of three
sparsely-acting labels [Al/H], [S/H] and [K/H] are shown for two spectral
regions.  The locations of known strong atomic lines are indicated by vertical
markers \citep{Smith_2013}. Although no information about atomic line lists
enter into our spectral model, the spectral derivatives for [Al/H] and [K/H] 
show strong contributions at wavelengths of known atomic lines.

\begin{figure}[p]
\centering
\includegraphics[width=\textwidth]{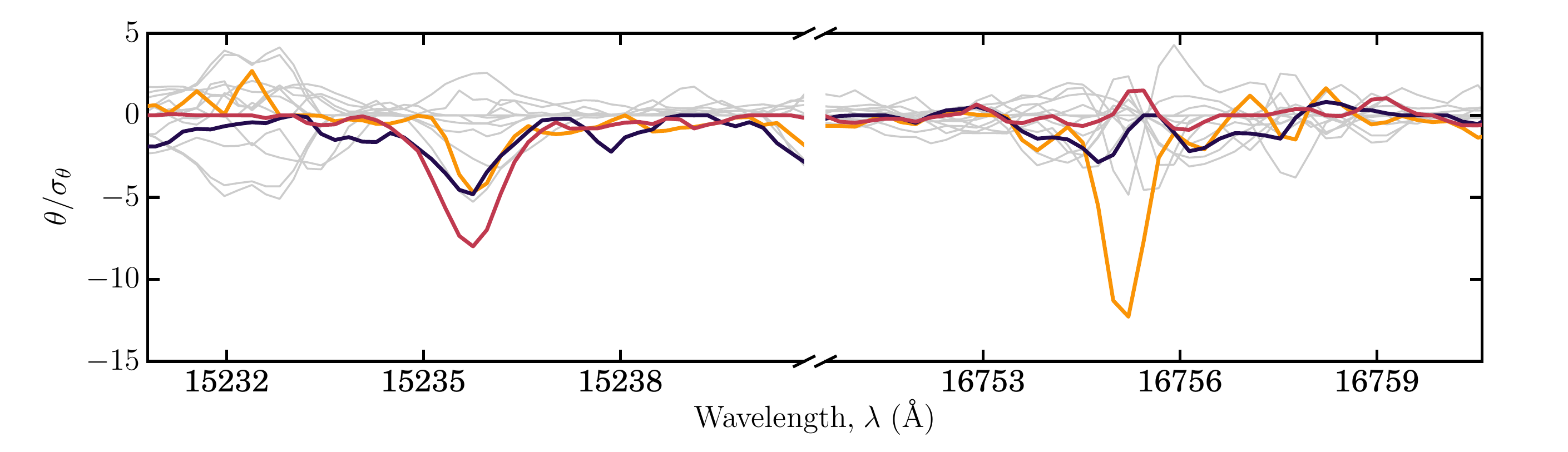}
\caption{First-order spectral derivatives (normalized to the standard deviation for that derivative) for all abundance labels centered at the \apogee\ `unknown' lines at 15235.8~\AA\ and 16755.6~\AA. Abundance derivatives are shown in grey, or colored as per Figure \ref{fig:line-identification}.\label{fig:line-identification-2}}
\end{figure}

The two unmarked spectral lines in Figure \ref{fig:line-identification} correspond
to atomic lines that are seen in \apogee\ spectra, but for which there is
currently no atomic data available.  The element that
is responsible for these spectral lines is unknown \citep{Shetrone_2015}.
Figure \ref{fig:line-identification-2} shows a zoom-in around these
two regions, where we show \emph{all} first-order spectral derivatives and
color [S/H] and [Al/H] as per Figure \ref{fig:line-identification}.  While
there is considerable blending of spectral derivatives near the line at 15235~\AA{}, [S/H] shows
the dominant first-order spectral coefficient at all pixels that include
this (previously unknown) atomic line.  For the previously unknown atomic line
at 16755~\AA{}, the spectral derivatives in our regularized model strongly
suggest that this is an Al transition. At 16755~\AA{} the [Al/H] spectral derivative
is nearly the strongest seen anywhere in the \apogee\ spectra, matched only
by two nearby (known) Al lines, and there is no significant blending by
spectral derivatives of other abundance labels.

We made no effort to cross-match our model with the list of ``unknown'' spectral lines
identified by the \apogee\ group.  This example presented itself
serendipitously whilst examining the relative strengths of different
first-order derivatives.  Indeed, the list of ``unknown spectral lines''
was unknown to the lead-author when Figure \ref{fig:line-identification} was first produced.  While it appears
likely that the two unknown lines shown here arise from S and Al, respectively,
we stress that the spectral derivatives should only be used as a guide.  The
abundances of different elements are strongly covariant in nature (see below).
For this reason the spectral derivatives are merely indicative as to
which element label (of the set that we have available) correlates
most with the fluxes at the given pixels.  We have not performed
a detailed examination to understand \emph{how interpretable} this
\emph{Cannon} model is, and what physical insights it can provide. 
Convincing examples of this covariant behaviour (and methods to address
it) will be presented in a companion paper (Ness et al., 2016, in preparation).

Indeed, there is considerable room for improvement over the work
we have presented here.  For example, at least one of our model 
assumptions is provably false. \apogee\ spectra have
different resolutions.  The line spread function is both wavelength- and 
fibre-dependent.  Given this information, our current implementation of \TheCannon\ is  
sub-optimal: we assume nothing about differing line spread functions between 
stars in the reference or test set.  Similarly, we make no effort to 
accommodate fast-rotating stars, where the effect on the spectrum is 
approximately represented by convolution with a Gaussian kernel (in the same way
a lower resolution would be).  Tests performed by M.~Ness have revealed that 
this information is present in the data: if the fibre number is included as a
label in $\ell$, then the fibre number can be accurately inferred from stars
in the test set!  This is possible because the 
\apogee\ resolution varies smoothly (to some degree) with fibre number.  For 
these reasons it is clear that simultaneously solving for additional rotation
at test time would be an useful extension of this work.

Notwithstanding these issues, \TheCannon\ clearly has its place amongst
the current methods employed to analyse stellar spectra.  Precise labels
are returned even in the presence of noisy data, and the model internals
are sufficiently interpretable that they can inform physics-based models.
Admittedly, our comparisons to the globular cluster literature suggests
that our [Fe/H] (and other?) labels may be inaccurate in the metal-poor
regime, a signature arising from incompleteness and inaccuracy of our
training set labels.  However we have demonstrated that this limitation does not 
hinder any potential inferences using individual chemical abundances:
for these (and many other purposes) we care more about abundance precision,
not accuracy. With increased precision, we have shown that globular clusters
exhibit significantly narrower intrinsic spreads in abundances (and their anti-correlations)
than previously considered.  Crucially, because the detailed chemical
abundances reflect the star forming conditions and subsequent evolution of
the cluster, this work is a necessary and fundamental step forward in
understanding the evolution of stars, clusters, and ultimately, the Milky Way.

All of the code for this project is available with documentation at \url{http://thecannon.io/}.  
We encourage open community discussion by way of GitHub issues, and code contributions through pull requests.

\acknowledgements
The authors warmly thank Daniel Foreman-Mackey (University of Washington),
						 Jason Sanders (Cambridge), and
						 Angus Williams (Cambridge) for valuable discussions.
						 
This project was funded in part by
  the European Research Council under the European Union's Seventh Framework 
  Programme (FP~7) \acronym{ERC} Grant Agreement 320360,
  the \acronym{NSF} (grants \acronym{IIS-1124794}, \acronym{AST-1517237}),
  \acronym{NASA} (grant \acronym{NNX12AI50G}), and 
  the Moore-Sloan Data Science Environment at \acronym{NYU}.
This research made use of: 
  the \acronym{NASA} \project{Astrophysics Data System Bibliographic Services},
  \texttt{TOPCAT} \citep{Taylor_2005},
  GitHub, 
  Travis CI, the
  \texttt{IPython} \citep{Perez_2007}, 
  \texttt{numpy} \citep{van_der_Walt_2011}, 
  \texttt{matplotlib} \citep{Hunter_2007}, and
  \texttt{scipy} \citep{Jones_2001} packages, as well as
  Astropy, a community-developed core Python package for Astronomy \citep{astropy}.

This project made use of \sdssiii\ data.
Funding for \sdssiii\ has been provided by the Alfred P. Sloan
Foundation, the Participating Institutions, the National Science
Foundation, and the \acronym{U.S.} Department of Energy Office of Science. The
\sdssiii\ web site is http://www.sdss3.org/.

\sdssiii\ is managed by the Astrophysical Research Consortium for the
Participating Institutions of the \sdssiii\ Collaboration including the
University of Arizona, the Brazilian Participation Group, Brookhaven
National Laboratory, Carnegie Mellon University, University of
Florida, the French Participation Group, the German Participation
Group, Harvard University, the Instituto de Astrofisica de Canarias,
the Michigan State/Notre Dame/\acronym{JINA} Participation Group, Johns Hopkins
University, Lawrence Berkeley National Laboratory, Max Planck
Institute for Astrophysics, Max Planck Institute for Extraterrestrial
Physics, New Mexico State University, New York University, Ohio State
University, Pennsylvania State University, University of Portsmouth,
Princeton University, the Spanish Participation Group, University of
Tokyo, University of Utah, Vanderbilt University, University of
Virginia, University of Washington, and Yale University.

\clearpage

\clearpage

\clearpage

\clearpage

\clearpage

\clearpage

\clearpage

\clearpage

\clearpage

\clearpage

\clearpage

\clearpage

\clearpage

\clearpage

\clearpage

\clearpage

\clearpage

\begin{table*}
\centering
\begin{tabular}{lr}
\hline
Label & Error floor \\
\hline
$\Teff$ 		& 22~K \\
$\logg$ 		& 0.03~dex \\
$[\rm{C/H}]$ 	& 0.03~dex \\
$[\rm{N/H}]$  	& 0.04~dex \\
$[\rm{O/H}]$  	& 0.03~dex \\
$[\rm{Na/H}]$ 	& 0.08~dex \\
$[\rm{Mg/H}]$ 	& 0.03~dex \\
$[\rm{Al/H}]$ 	& 0.06~dex \\
$[\rm{Si/H}]$ 	& 0.03~dex \\
$[\rm{S/H}]$	& 0.05~dex \\
$[\rm{K/H}]$ 	& 0.07~dex \\
$[\rm{Ca/H}]$ 	& 0.05~dex \\
$[\rm{Ti/H}]$ 	& 0.07~dex \\
$[\rm{V/H}]$ 	& 0.16~dex \\
$[\rm{Mn/H}]$ 	& 0.04~dex \\
$[\rm{Fe/H}]$ 	& 0.01~dex \\
$[\rm{Ni/H}]$ 	& 0.03~dex \\
\hline
\end{tabular}
\caption{Minimum error floor in labels, taken as the error in estimating labels for validation set stars at S/N $= 50$ (Figure \ref{fig:label-recovery-snr}).}
\label{tab:error-floors}
\end{table*}

\clearpage

\begin{table*}
\centering
\begin{tabular}{lccccc}
\hline
\texttt{APOGEE\_ID} & $\Teff$ & $\logg$ & [Al/H] & [Ca/H] & $\cdots$ \\
& (K) & & & \\
\hline
2M18482322+4727598 & $4502 \pm 3$ & $2.279 \pm 0.008$ & $\pn0.275 \pm 0.013$ & $-0.019   \pm 0.012$ & $\cdots$ \\
2M18495146+4622483 & $4592 \pm 5$ & $2.641 \pm 0.010$ & $\pn0.385 \pm 0.016$ & $\pn0.109 \pm 0.018$ & $\cdots$ \\
2M18543638+4658349 & $4767 \pm 7$ & $2.550 \pm 0.014$ & $-0.076   \pm 0.019$ & $-0.307   \pm 0.017$ & $\cdots$ \\
2M18471131+4726561 & $4494 \pm 3$ & $2.158 \pm 0.009$ & $\pn0.040 \pm 0.014$ & $-0.187   \pm 0.013$ & $\cdots$ \\
2M18493578+4838112 & $4720 \pm 6$ & $2.828 \pm 0.011$ & $\pn0.055 \pm 0.019$ & $-0.074   \pm 0.017$ & $\cdots$ \\
2M18442845+4651265 & $4659 \pm 5$ & $2.559 \pm 0.010$ & $\pn0.042 \pm 0.014$ & $-0.046   \pm 0.013$ & $\cdots$ \\
2M18464037+4816449 & $4730 \pm 4$ & $2.621 \pm 0.009$ & $\pn0.372 \pm 0.013$ & $\pn0.204 \pm 0.013$ & $\cdots$ \\
2M18515439+4809330 & $4786 \pm 6$ & $2.601 \pm 0.013$ & $\pn0.009 \pm 0.017$ & $-0.208   \pm 0.016$ & $\cdots$ \\
\hline
\end{tabular}
\caption{All 17-labels for 87,563 \apogee\ giant stars using the regularized model. Positions and ancillary measurements from \aspcap\ are included for convenience. Only formal errors are listed here. See text for full details regarding the known error budget.  This table is online-only. A portion is shown here to indicate form and content.}
\label{tab:the-good-stuff}
\end{table*}


\begin{thebibliography}{dummy}\raggedright

\bibitem[Alam et al.(2015)]{Alam_2015} Alam, S., Albareti, F.~D., 
Allende Prieto, C., et al.\ 2015, \apjs, 219, 12 

\bibitem[Astropy Collaboration et 
al.(2013)]{astropy} Astropy Collaboration, Robitaille, T.~P., Tollerud, E.~J., et al.\ 2013, Astronomy \& Astrophysics, 558, AA33 

\bibitem[Cannon 
\& Pickering(1912)]{Cannon_1912} Cannon, A.~J., \& Pickering, E.~C.\ 1912, Annals of Harvard College Observatory, 56, 115 

\bibitem[Cohen et al.(2005)]{Cohen_2005} Cohen, J.~G., Briley, 
M.~M., \& Stetson, P.~B.\ 2005, \aj, 130, 1177 

\bibitem[De Silva et al.(2015)]{De_Silva_2015} De Silva, G.~M., 
Freeman, K.~C., Bland-Hawthorn, J., et al.\ 2015, \mnras, 449, 2604 

\bibitem[Garc{\'{\i}}a P{\'e}rez et al.(2015)]{Garcia_Perez_2015} 
Garc{\'{\i}}a P{\'e}rez, A.~E., Allende Prieto, C., Holtzman, J.~A., et 
al.\ 2015, arXiv:1510.07635 

\bibitem[Gilmore et al.(2012)]{Gilmore_2012} Gilmore, G., Randich, 
S., Asplund, M., et al.\ 2012, The Messenger, 147, 25 

\bibitem[Harris(1996)]{Harris_1996} Harris, W.~E.\ 1996, \aj, 112, 
1487 

\bibitem[Ho et al.(2016)]{Ho_2016} Ho, A.~Y.~Q., Ness, M.~K., 
Hogg, D.~W., et al.\ 2016, arXiv:1602.00303 

\bibitem[Hogg et al.(2016)]{Hogg_2016} Hogg, D.~W., Casey, A.~R., 
Ness, M., Rix, H.-W., \& Foreman-Mackey, D.\ 2016, arXiv:1601.05413 

\bibitem[Holtzman et al.(2015)]{Holtzman_2015} Holtzman, J.~A., 
Shetrone, M., Johnson, J.~A., et al.\ 2015, \aj, 150, 148 

\bibitem[Hunter(2007)]{Hunter_2007} Hunter, J.~D.\ 2007, Matplotlib: A 2D Graphics Environment, Computing in Science \& Engineering, 9, 90-95

\bibitem[Jones et al.(2001)]{Jones_2001} Jones, E., Oliphant E., Peterson P., et al.\ 2001, SciPy: Open Source Scientific Tools for Python, http://www.scipy.org/ (Online; accessed 2016-02-16)

\bibitem[M{\'e}sz{\'a}ros et al.(2015)]{Meszaros_2015} 
M{\'e}sz{\'a}ros, S., Martell, S.~L., Shetrone, M., et al.\ 2015, \aj, 149, 
153 
\bibitem[Ness et al.(2015a)]{tc} Ness, M., Hogg, D.~W., 
Rix, H.-W., Ho, A.~Y.~Q., \& Zasowski, G.\ 2015, \apj, 808, 16

\bibitem[Ness et al.(2015b)]{age} Ness, M., Hogg, D.~W., 
Rix, H., et al.\ 2015, arXiv:1511.08204 

\bibitem[Perez \& Granger(2007)]{Perez_2007} P\'erez, F., \& Granger, B. E.\ 2007, IPython: A System for Interactive Scientific Computing, Computing in Science \& Engineering, 9, 21-29

\bibitem[Shetrone et al.(2015)]{Shetrone_2015} Shetrone, M., Bizyaev, 
D., Lawler, J.~E., et al.\ 2015, \apjs, 221, 24 

\bibitem[Smith et al.(2013)]{Smith_2013} Smith, V.~V., Cunha, K., 
Shetrone, M.~D., et al.\ 2013, \apj, 765, 16 

\bibitem[Taylor(2005)]{Taylor_2005} Taylor, M.~B.\ 2005, Astronomical Data Analysis Software and Systems XIV, 347, 29 

\bibitem[Tibshirani(1996)]{Tibshirani_1996} Tibshirani, R.\ 1996, Journal of the Royal Statistical Society. Series B (Methodological), 58, 267-288

\bibitem[van der Walt et al.(2011)]{van_der_Walt_2011} van der Walt, S., Colbert S.~C., Varoquaux, G.\ 2011, The NumPy Array: A Structure for Efficient Numerical Computation, Computing in Science \& Engineering, 13, 22-30

\bibitem[Zasowski et al.(2013)]{Zasowski_2013} Zasowski, G., Johnson, 
J.~A., Frinchaboy, P.~M., et al.\ 2013, \aj, 146, 81 


\end{thebibliography}
\end{document}